\documentclass[10pt,oneside,final]{IEEEtran}

\usepackage{amssymb}
\usepackage{amsmath}

\usepackage{graphicx}
\usepackage{color}
\usepackage{multicol}

\usepackage{bm}
\usepackage{latexsym}
\usepackage{stfloats}
\usepackage{float}
\usepackage{subfigure}
\usepackage{exscale}
\usepackage{relsize}
\usepackage{cite}
\usepackage{cases}
\usepackage{algorithm}
\usepackage{algorithmicx}
\usepackage{algpseudocode}

\usepackage{epsfig}
\usepackage{epstopdf}
\usepackage{breqn}
\usepackage{enumerate}
\usepackage{multirow}
\usepackage[utf8]{inputenc}

\usepackage{mathtools}

\usepackage{setspace}
\usepackage{url}
\usepackage{makecell}
\usepackage{diagbox}

\usepackage{xcolor}
\usepackage{xpatch}

\usepackage[colorlinks,
linkcolor=blue,
urlcolor=black,
anchorcolor=blue,
citecolor=blue]{hyperref}
\pdfoutput=1

\algrenewcommand{\algorithmicrequire}{\textbf{Input:}}  
\algrenewcommand{\algorithmicensure}{\textbf{Output:}} 

\begin{document}

\title{\color{blue}Target Localization in Cooperative ISAC Systems: A Scheme Based on 5G NR OFDM Signals}

\author{
	Zhenkun Zhang, Hong Ren, Cunhua Pan, Sheng Hong, Dongming Wang, Jiangzhou Wang, \textit{Fellow, IEEE},
	and Xiaohu You, \textit{Fellow, IEEE}
	\thanks{
		Zhenkun Zhang, Hong Ren, Cunhua Pan, Dongming Wang, Jiangzhou Wang and Xiaohu You are with National Mobile Communications Research Laboratory, Southeast University, Nanjing 210096, China (e-mail:{zhenkun\_zhang, hren, cpan, wangdm, j.z.wang, xhyu}@seu.edu.cn).
		
		Sheng Hong is with the Information Engineering School, Nanchang University, Nanchang, Jiangxi 330031, China (e-mail: shenghong@ncu.edu.cn).}
	\vspace{-0.4cm}
	}
%

%
%
\maketitle

\newtheorem{lemma}{Lemma}
\newtheorem{theorem}{Theorem}
\newtheorem{remark}{Remark}
\newtheorem{corollary}{Corollary}
\newtheorem{proposition}{Proposition}

%
\begin{abstract}
The integration of sensing capabilities into communication systems, by sharing physical resources, has a significant potential for reducing spectrum, hardware, and energy costs while inspiring innovative applications. 
Cooperative networks, in particular, are expected to enhance sensing services by enlarging the coverage area and enriching sensing measurements, thus improving the service availability and accuracy.
This paper proposes a cooperative integrated sensing and communication (ISAC) framework by leveraging information-bearing orthogonal frequency division multiplexing (OFDM) signals transmitted by access points (APs). 
Specifically, we propose a two-stage scheme for target localization, where communication signals are reused as sensing reference signals based on the system information shared at the central processing unit (CPU).
{\color{blue}
In Stage I, we propose a two-dimensional fast Fourier transform (2D-FFT)-based algorithm to measure the ranges of scattered paths induced by targets, through the extraction of delay and Doppler information from the sensing channels between APs. 
Then, the target locations are estimated in Stage II based on these range measurements.
Considering the potential occurrence of ill-conditioned measurements with large error during the extraction of time-frequency information, we propose an efficient algorithm to match the range measurements with the targets while eliminating ill-conditioned measurements, achieving high-accuracy target localization.
In addition, based on the transmission configurations defined in the fifth generation (5G) standards, we elucidate the performance trade-offs in both communication and sensing, and extend the proposed sensing scheme for general scenarios.}
Finally, numerical results confirm the effectiveness of our sensing scheme and the cooperative gain of the ISAC framework.


\end{abstract}

%

%

%
\section{Introduction}
Driven by the emerging services requiring high-level communication and sensing capabilities, e.g., vehicle to everything (V2X), extended reality and smart home, integrated sensing and communication (ISAC) has been envisioned as a pivotal technology for the sixth generation (6G) wireless networks \cite{9625159}.
By sharing core hardware, space-time-frequency resources and/or transmit signals within a single system in a symbiotic manner, ISAC networks are able to improve the spectral and energy efficiencies, while reducing the hardware scale and costs \cite{8999605}.
In addition, mutual enhancement for both communication and sensing performances is expected to be achieved by the coexistence of the two functionalities.
For example, the transmission rate and robustness of wireless communication systems can be enhanced with the knowledge of propagation environment \cite{6924849}.

Depending on the employed waveform, ISAC schemes can be classified into two primary categories: sensing-centric and communication-centric.
Sensing-centric schemes manipulate the radar signals, enabling the modulation of a limited number of bits \cite{8828023,7944444}.
Nevertheless, the communication capabilities of these sensing-centric ISAC systems are extremely constrained since the signal autocorrelation, which is critical to radar sensing, is reduced when modulated by high-order random symbols.
Hence, they are unable to meet the data throughput demands of various ISAC applications.
In contrast, the communication-centric ISAC schemes focus on the implementation of sensing functionality based on the communication signals.
Then, the use case of the communication system can be largely expanded beyond data transmissions, realizing applications such as localization and tracking, environment sensing, gesture and activity recognition, which can also be triggered for potential novel applications \cite{9376324,9737357}.
An intuitive advantage of communications-centric ISAC scheme is its potential to leverage the widespread deployment of numerous communication nodes, such as base stations (BSs), WiFi routers and user equipments (UEs), for sensing \cite{9516898}. 
This greatly reduces the difficulty and cost of ISAC system deployment, especially in indoor and complex urban scenarios.
Moreover, the communication network-based ISAC system can benefit from the cooperation between the communication nodes \cite{5762797}.
Thanks to the wider distribution area and/or dense deployment of communication nodes, the cooperative ISAC networks host abundant sources of sensing measurements and computing capability \cite{10049304,7482747}, enabling high performance in terms of coverage range, sensing accuracy, robustness, flexibility, and so on \cite{9814522}.

For the UEs equipped with signal transmission capabilities and registered within communication networks, their locations can be efficiently estimated by employing the established \textit{device-based} sensing methodologies \cite{8226757}.
These methodologies involve the extraction of sensing measurements, such as time-of-arrival and angle-of-arrival, from dedicated reference signals exchanged between UEs and BSs.
However, for passive targets not interconnected within the communication networks, the adoption of a \textit{device-free} sensing approach becomes imperative. 
In a similar manner to radar systems, device-free sensing schemes detect, localize and/or track targets solely by analyzing the signals reflected by the target contained within the communication signals arrived at the receivers.
This thus motivates our work on the development of novel methods for device-free sensing.

The primary obstacle in implementing device-free sensing within communication networks arises from the unfavourable radar sensing characteristics inherent in the random waveforms required for data transmission \cite{10004202}.
Notably, orthogonal frequency division multiplexing (OFDM) \cite{5281762,6094142} is the predominant waveform for 5G cellular networks and a prominent candidate waveform for 6G.
To address this issue, several studies have explored device-free sensing methods for OFDM-based cellular networks.
A sensing framework was designed in \cite{2023arXiv231101674L}, where the BS searches candidate sensing targets using a rotating beam to scan the service area.
Based on the echoes of multiple input multiple-output (MIMO)-OFDM communication signals with \textit{cyclic prefix} (CP), a preliminary two-stage framework for sensing measurement extraction and multi-target localization was proposed in \cite{9367457}. 
By utilizing the positioning reference signal (PRS) in 5G new radio (NR) standards as a sensing reference signal, \cite{9921271} proposed a sensing scheme for passive targets and analyzed its performance for range and velocity estimation.
In \cite{8062243}, a signal-stripping-based approach for sensing parameter estimation was designed, leveraging estimated data symbols and channels.
However, the aforementioned studies only focused on the sensing approaches for each communication node, without considering the cooperative gain on sensing performance of the entire ISAC system.

Utilizing networked nodes in communication systems, cooperative sensing is expected to satisfy the need of practical applications for large coverage, precise and robust sensing \cite{10273396}.
Therefore, the cooperative interactions among communication nodes has been explored in a few contributions from the following aspects.
\begin{enumerate}
	\item 
	\textit{Cooperation framework: } 
	{\color{blue}
		In \cite{9909950}, a frame structure design was proposed to facilitate the sensing applications among BSs in adjacent communication cells. 
		Recently, the authors of \cite{hanetalcellularnet} designed feasible topology and frame structure to enable sensing cells in mobile communication networks at a low cost.
		In addition, a two-BS cooperative sensing scheme for perceptive mobile networks was proposed in \cite{10616023}, where each sensing BS jointly utilizes the echoes of sensing signals transmitted by both BSs to estimate the locations and velocities of targets.
		Note that the synchronization parameters, extracted sensing measurements, and exchanged signals between nodes can serve as prior information for target sensing.
		Information sharing and data fusion are highly necessary for cooperative sensing tasks, and potentially enable the sensing schemes based on information-bearing communication signals.
	}
	Thus, the networks with centralized signal processing architecture are ideal platforms for the ISAC systems.
	Based on the cloud radio access network (C-RAN) architecture \cite{8344451}, the authors of \cite{8827589} proposed a scheme for sensing parameter extraction and clutter reduction by developing a general signal model for both uplink and downlink orthogonal frequency-division multiple access (OFDMA) MIMO signals.

	\item 
	\textit{Performance analysis and transmission design: } 
	Investigating a rate-splitting scheme for C-RAN-based ISAC systems, the authors of \cite{10032141} optimized communication and sensing performance by analyzing the Pareto boundary of the performance region. 
	Cell-free MIMO network, in which multiple distributed access points (APs) are connected to a central processing unit (CPU) via high- capacity fronthaul links, emerged as a scalable version of C-RAN \cite{8630677}.
	Recently, the authors of \cite{10207026} studied the cell-free massive MIMO ISAC system, focusing on the optimization of user scheduling and power allocation.

	\item 
	\textit{Sensing algorithm design: } 
	In \cite{9724258,10001615,10226276}, the algorithms for joint data association and target localization were designed for the corresponding cooperative sensing schemes.
	By extending the ISAC framework in \cite{9367457} to a multi-BS cellular network, the scheme in \cite{9724258} estimated target locations using range measurements obtained by the BSs.
	The authors of \cite{10001615} proposed a trilateration sensing scheme for a two-BS network with the assistance of a reconfigurable intelligent surface.
	For the localization and velocity estimation of a moving target, the scheme in	\cite{10226276} fused the phase features that are extracted from the demodulation symbols of BSs.
	
\end{enumerate}
 
{\color{blue}
	However, the existing works has the following limitations:
	Firstly, in the sensing schemes in \cite{10616023,9724258,10001615,10226276}, each node is required to transmit signals and receive echoes in a full-duplex manner, leading to self-interference that cannot be completely mitigated using the existing methods.
	Secondly, unlike the cooperative sensing scheme in \cite{9724258} that uses only the echo of the signal transmitted by the sensing node itself, the schemes in \cite{hanetalcellularnet} and \cite{10616023} can leverage signals transmitted by other nodes to extract additional sensing information, while the accuracy of extracted information is significantly affected by the quality of synchronization among the nodes.
	To mitigate the impacts of symbol timing offset (STO) and carrier frequency offset (CFO), a method was proposed in \cite{10616023} for the scenario with no line-of-sight (LoS) path between the nodes.
	Nonetheless, the strategies for effectively utilizing the prevalent LoS paths in general scenarios to address the effects of imperfect synchronization remain under-explored in the existing ISAC schemes.
	Finally, the design of algorithms for target sensing, particularly regarding data fusion, is a key challenge in the implementation of ISAC systems.
	The sensing schemes discussed in \cite{9724258,10001615,10226276} assumed that every target is correctly detected by all cooperative nodes, which limits the robustness of the algorithms in the presence of ill-conditioned measurements with extremely large errors.	

In this paper, we focus on the design of practical cooperative ISAC framework that are compatible with the 5G NR standards.}
We consider a cell-free MIMO network as the cooperation architecture, where multiple APs and UEs exchange communication signals, while the received downlink signals are reused as sensing reference signals for target localization.
This design minimizes the modification to the existing wireless communication systems, without requiring full-duplex operation of the APs.
To achieve the device-free sensing based on the practical OFDM communication signals, effective signal processing method is proposed to extract range measurements and estimate the locations of targets.
The main contributions of this paper are summarized as follows:
\begin{enumerate}
	\item
	Firstly, we propose a novel cooperative ISAC framework based on a cell-free network consisting of multiple communication APs. 
	Inspired by the frame structure employed in wireless communications, the framework facilitates device-free target localization without relying on dedicated sensing reference signals.
	Based on the comprehensive information sharing at the CPU, our proposed scheme achieves target sensing solely utilizing the received information-bearing OFDM signals that are transmitted from other APs and reflected by the targets without sacrificing communication performance.
	
	\item
	Secondly, we propose an efficient two-stage scheme for cooperative device-free target localization.
	We consider the signal model of the commonly used OFDM symbol with CP.
	Note that the passive targets introduce scattered paths among the APs. 	
	{\color{blue}
	In the first stage, based on the received communication signals, a two-dimensional fast Fourier transform (2D-FFT)-based algorithm is developed to extract the delay and Doppler information of the paths between the APs.
	By regarding the LoS paths as reference paths, we estimate and compensate for the STOs and CFOs to obtain the length of scattered paths, which are equivalent to \textit{bistatic range measurements} in radar systems.
	In the second stage, we propose an algorithm based on the maximum-likelihood (ML) method to jointly associate each range measurement with target and estimate the target locations.
	In particular, the proposed sensing scheme is capable of eliminating the ill-conditioned measurements in the estimated bistatic ranges.}
	
	\item
	{\color{blue}
	Thirdly, we discuss the performance trade-offs achieved in the proposed cell-free cooperative ISAC scheme based on the transmission configurations specified in the Third Generation Partnership Project (3GPP) standards.
	Furthermore, the sensing scheme is modified for the general scenarios, where more flexible performance trade-offs and higher sensing availability for the system can be achieved.}

	\item 
	Finally, based on the extensive simulation results, we validate the effectiveness of the proposed cooperative ISAC framework and sensing scheme.
	Insights about the performance of ISAC systems are also derived from the simulation results, providing guidance for engineering practice.
	
\end{enumerate}

The remainder of this paper is organized as follows.
Section \ref{Sec_frame_struc} introduces the frame structure of 5G.
Section \ref{Sec_sys_mod} describes the proposed cooperative ISAC framework and the signal model.
In Section \ref{Sec_sens_scheme}, we develop a sensing scheme to extract range measurements and estimate the locations of targets.
In Section \ref{Sec_tradeoff_extension}, the trade-offs in practical cooperative ISAC systems are analyzed and the sensing scheme is modified for general scenarios.
Extensive simulation results and discussions are given in Section \ref{Sec_simu_resul}.
Finally, Section \ref{Sec_conclusion} draws the conclusions of this work.

\emph{Notations}:
${\mathbb C}$ and $j=\sqrt {{\rm{ - 1}}}$ denote the complex field and the imaginary unit, respectively.
Vectors, matrices and sets are represented as boldface lowercase, boldface uppercase and calligraphic uppercase letters, respectively.
For a matrix ${\mathbf A}$, its transpose and Hermitian are denoted by ${\bf A}^{\mathrm T}$, ${\bf A}^{\mathrm H}$, respectively.
For a vector $ {\mathbf{a}} $, $\left\| {\mathbf{a}} \right\|_1$ and $\left\| {\mathbf{a}} \right\|_2$ respectively denote its $l_1$- and $l_2$-norm, and $\mathrm{diag}\left( {\mathbf{a}}  \right)$ denotes a diagonal matrix where the elements of $ {\mathbf{a}} $ form the main diagonal elements of the matrix.
For a set $\mathcal{A}$, $\left| \mathcal{A} \right|$ denotes the number of elements in $\mathcal{A}$.
{\color{blue}
	$\mathbf{A}\odot \mathbf{B}$ denotes the Hadamard product of matrix $\mathbf{A}$ and $\mathbf{B}$.}

\renewcommand{\arraystretch}{1.2}
\begin{table}[tb]
	\begin{center}
		\caption{Supported OFDM Numerologies and Corresponding FRs \cite{3GPP138211}}
		\label{Table_FR_CP}
		\begin{tabular}{|c|c|c|c|}
			\hline
			$\mu$ & CP               & FR(s)   & $N_{\mathrm{f}}$\\
			\hline
			0, 1  & Normal           & FR1    & \multirow{3}{*}{4096}\\
			\cline{1-3} 
			2     & Normal, Extended & FR1, FR2 &\\
			\cline{1-3} 
			3 - 6 & Normal           & FR2    &\\
			\hline
		\end{tabular}
	\end{center}
\vspace{-0.2cm}
\end{table}

\section{5G NR Frame Structure}\label{Sec_frame_struc}
In this section, we review the frame structure specified in 3GPP standard for 5G NR, focusing on the resource allocations between CP and data symbols.
\subsection{Numerologies}
An OFDM symbol is the smallest time-domain resource unit in 5G NR, and a CP is inserted at the beginning of each OFDM symbol.
According to 3GPP TS 38.211 Release 17 \cite{3GPP138211}, the OFDM numerologies refer to the configurations of subcarrier spacing (SCS) and CP.
Specifically, there are seven supported configurations of SCS $\Delta f^\mu$ given as follows:
\begin{align}\label{Delta_f}
	\Delta f^\mu = 2^\mu \times 15 \;\text{[KHz]},
\end{align}
where $\mu \in \left\{ 0,1,\dots ,6 \right\} $ is the configuration parameter,
while the optional CP configurations include two modes: \textit{normal} and \textit{extended}.

The frequency bands available to NR are divided into two frequency ranges (FRs), namely FR1 (0.41 GHz - 7.125 GHz) and FR2 (24.25 GHz - 52.6 GHz) \cite{3GPP13810101,3GPP13810102}.
The SCS configurations applicable to each FR are summarized in Table \ref{Table_FR_CP}.
In addition, the FFT size, which is also the number of samples in a time-domain OFDM symbol without CP, is fixed at $N_{\mathrm{f}}=4096$.
Hence, the sampling interval of SCS configuration $\mu$ is
\begin{align}\label{Ts}
	T_{\mathrm{s}}^\mu=\frac{1}{N_{\mathrm{f}}\Delta f^\mu\times 10^3}=\frac{1}{2^{\mu +12}\times 15\times 10^3} \;\text{[s]}.
\end{align}
The normal CP can be used with all SCS configurations, while the extended CP is dedicated to the configuration with 60 KHz SCS in large delay spread transmission.
The CP configurations affect the position and duration of the CP for each OFDM symbol, which will be detailed in the next subsection.

\subsection{Structure of Frames, Subframes and Slots}
\begin{figure}
	\centering
	\includegraphics[width=0.98\linewidth]{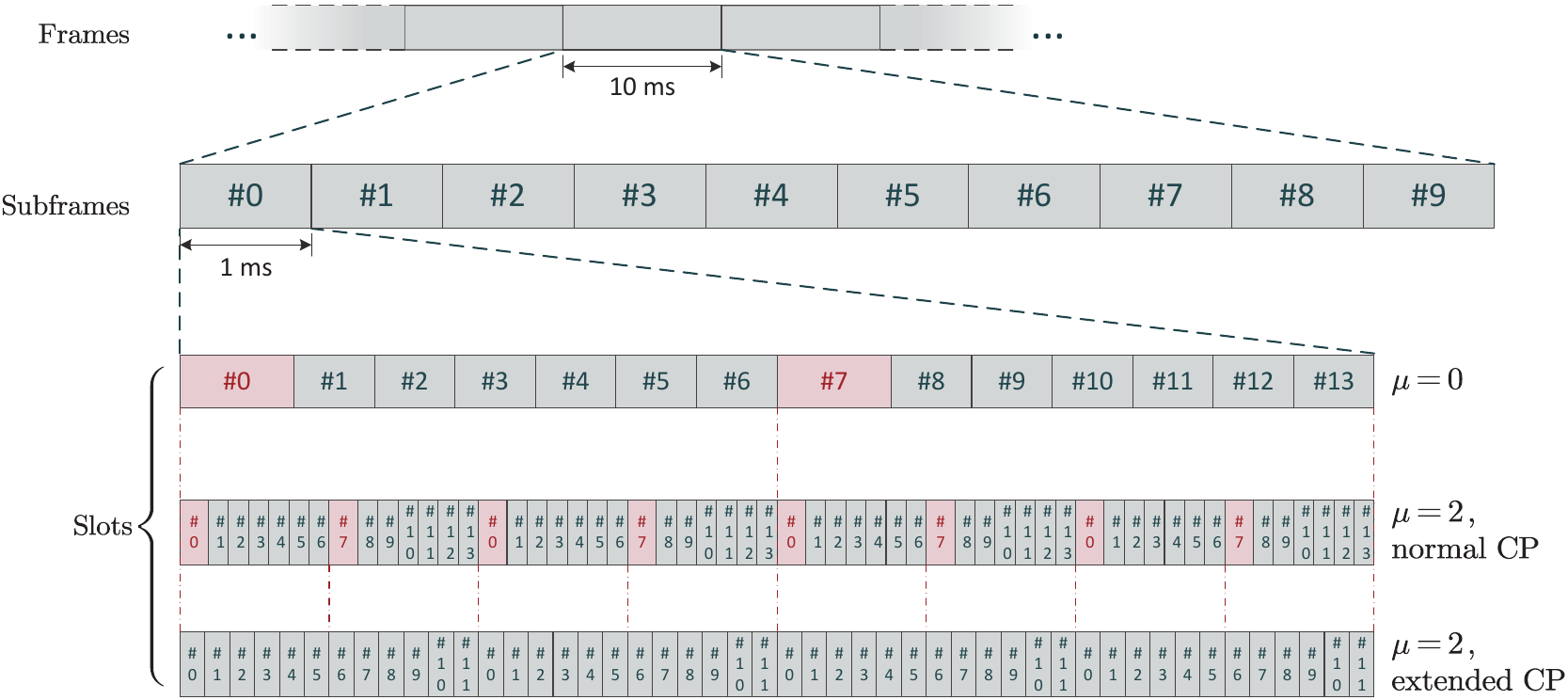}
	\caption{Transmission organization in 5G NR.}
	\label{FrameStructure}
	\vspace{-0.2cm}
\end{figure}
As illustrated in Fig. \ref{FrameStructure}, the transmission in NR is organized in frames with duration of 10 milliseconds (ms), each of which is divided into ten 1 ms-duration subframes.
Then, each subframe is equally divided into $2^\mu$ slots.
As shown in Table \ref{Table_slot}, each slot consists of $N_{\text{symb}}$ consecutive OFDM symbols numbered $l_{\mathrm s}\in \left\{ 0,\dots ,N_{\text{symb}}-1 \right\}$ in an increasing order with $N_{\text{symb}}=14$ for normal CP and $N_{\text{symb}}=12$ for extended CP.
Hence, the number of OFDM symbols in a frame is equal to $10\times 2^{\mu}\times N_{\mathrm{symb}}$.
From \eqref{Delta_f}, it can be observed that the number of transmitted OFDM symbols in a given duration increases with respect to (w.r.t.) SCS, resulting in a higher maximum data rate.

As specified in TS 38.211, the time unit is defined as $T_{\mathrm{c}}={{1}/{\left( 480\times 10^3N_{\mathrm{f}} \right)}}$ s.
For each configuration $\mu$, the duration of CP in terms of the number of time units is specified as follows \cite{3GPP138211}
\begin{align}\label{N_CP}
	N^{\mu,l_{\mathrm s}}_{\text{CP}}=\begin{cases}
		512\kappa \times 2^{-\mu} & \!\!\text{extended CP}\\
		144\kappa \times 2^{-\mu}+16\kappa &\!\! \text{normal CP}, \, l_{\mathrm s}\in\left\lbrace 0,7\right\rbrace \\
		144\kappa \times 2^{-\mu}&\!\! \text{normal CP}, \, l_{\mathrm s}\notin\left\lbrace 0,7\right\rbrace\\
	\end{cases},
\end{align}
where $\kappa =64$.
That is, the CP duration for each symbol in terms of time $T^{\mu,l_{\mathrm s}}_{\text{CP}}$ is given by $N^{\mu,l_{\mathrm s}}_{\text{CP}}T_{\mathrm{c}}$.
Meanwhile, the CP duration in terms of the number of sampling intervals $Q^{\mu,l_{\mathrm s}}_{\text{CP}}$ is shown in Table \ref{Table_slot}.
For simplicity, we will replace $Q^{\mu,l_{\mathrm s}}_{\text{CP}}$ with $Q$ in the remainder of this paper.

\begin{table*}[tb]
	\renewcommand{\arraystretch}{1.5}
	\begin{center}
		\caption{5G NR Structure of Frames, Subframes and Slots \cite{3GPP138211}}
		\label{Table_slot}
		\begin{tabular}{|c|c|c|c|c|}
			\hline
			CP  & $\mu$ &  \makecell[c]{Number of symbols per slot\\ $N_{\text{symb}}$}  & \makecell[c]{CP duration per symbol \\ $T^{\mu,l_{\mathrm s}}_{\text{CP}}$ [ms]} & \makecell[c]{Number of sampling points\\in CP per symbol $Q^{\mu,l_{\mathrm s}}_{\text{CP}}$ }\\
			\hline
			Normal   & 0 - 6 &  14  & 
			$\begin{cases}
				0.3\times 2^{-\mu -6}+1/1920\!\!&		l_{\mathrm{s}}\in \left. \left\{ 0,7 \right. \right\}\\
				0.3\times 2^{-\mu -6}&		\mathrm{others}\\
			\end{cases}$ & 
			$\begin{cases}
				288+ 2^{\mu+5}\!\!&l_{\mathrm s}\in\left\lbrace 0,7\right\rbrace\\
				288&\text{others}
			\end{cases}$ \\
			\hline
			Extended &  2    & 12  &			$4.17\times 10^{-3}$ &			1024 \\
			\hline
		\end{tabular}
	\end{center}
	\vspace{-0.2cm}
\end{table*}

{\color{blue}
\subsection{Transmission bandwidth configuration}
For the physical resource allocation of 5G NR, resource block (RB) constitutes the smallest resource unit in the frequency domain, each of which consists of 12 consecutive subcarriers. 
Based on the 3GPP standards \cite{3GPP13810101,3GPP13810102}, we summarize the maximum transmission bandwidth configuration $N_{\mathrm{RB}}$ for various channel bandwidths in Table \ref{Table_RB}.
It can be seen that due to the existence of guardband, the maximum bandwidth for transmission is narrower than the channel bandwidth.
For instance, under the configuration of $\mu =1$, a channel bandwidth of 50 MHz permits a maximum of $12\times 133=1596$ active subcarriers, resulting in a maximum transmission bandwidth of  $1596\times {\Delta f}^\mu=47.88$ MHz.
In the simulations of this work, we will set the number of active subcarriers according to Table \ref{Table_RB}.
}

\begin{table}[]\color{blue}
	\small
	\centering
	\caption{Maximum Transmission Bandwidth Configuration $N_{\mathrm{RB}}$ for Specific UE Channel Bandwidth Setups \cite{3GPP13810101,3GPP13810102}}
	\label{Table_RB}
		\begin{tabular}{|cc|c|c|c|c|}
		\hline
		\multicolumn{2}{|c|}{\begin{tabular}[c]{@{}c@{}}Channel\\ Bandwidth\end{tabular}} & 20 MHz & 50 MHz & 100 MHz & 200 MHz \\ \hline
		\multicolumn{1}{|c|}{\multirow{5}{*}{$\mu$}}               & 0                    & 106 RBs                                          & 270 RBs                                          & N/A                                               & N/A                                               \\ \cline{2-6} 
		\multicolumn{1}{|c|}{}                                     & 1                    & 51 RBs                                           & 133 RBs                                          & 273 RBs                                           & N/A                                               \\ \cline{2-6} 
		\multicolumn{1}{|c|}{}                                     & 2 (FR1)              & 24 RBs                                           & 65 RBs                                           & 135 RBs                                           & N/A                                               \\ \cline{2-6} 
		\multicolumn{1}{|c|}{}                                     & 2 (FR2)              & N/A                                              & 66 RBs                                           & 132 RBs                                           & 264 RBs                                           \\ \cline{2-6} 
		\multicolumn{1}{|c|}{}                                     & 3                    & N/A                                              & 32 RBs                                           & 66 RBs                                            & 132 RBs                                           \\ \hline
	\end{tabular}
\end{table}

\section{System Model}\label{Sec_sys_mod}
In this section, we propose an ISAC framework based on the cell-free MIMO networks.
The system architecture and signal model are presented.

\begin{figure}\color{blue}
	\centering
	\includegraphics[width=0.9\linewidth]{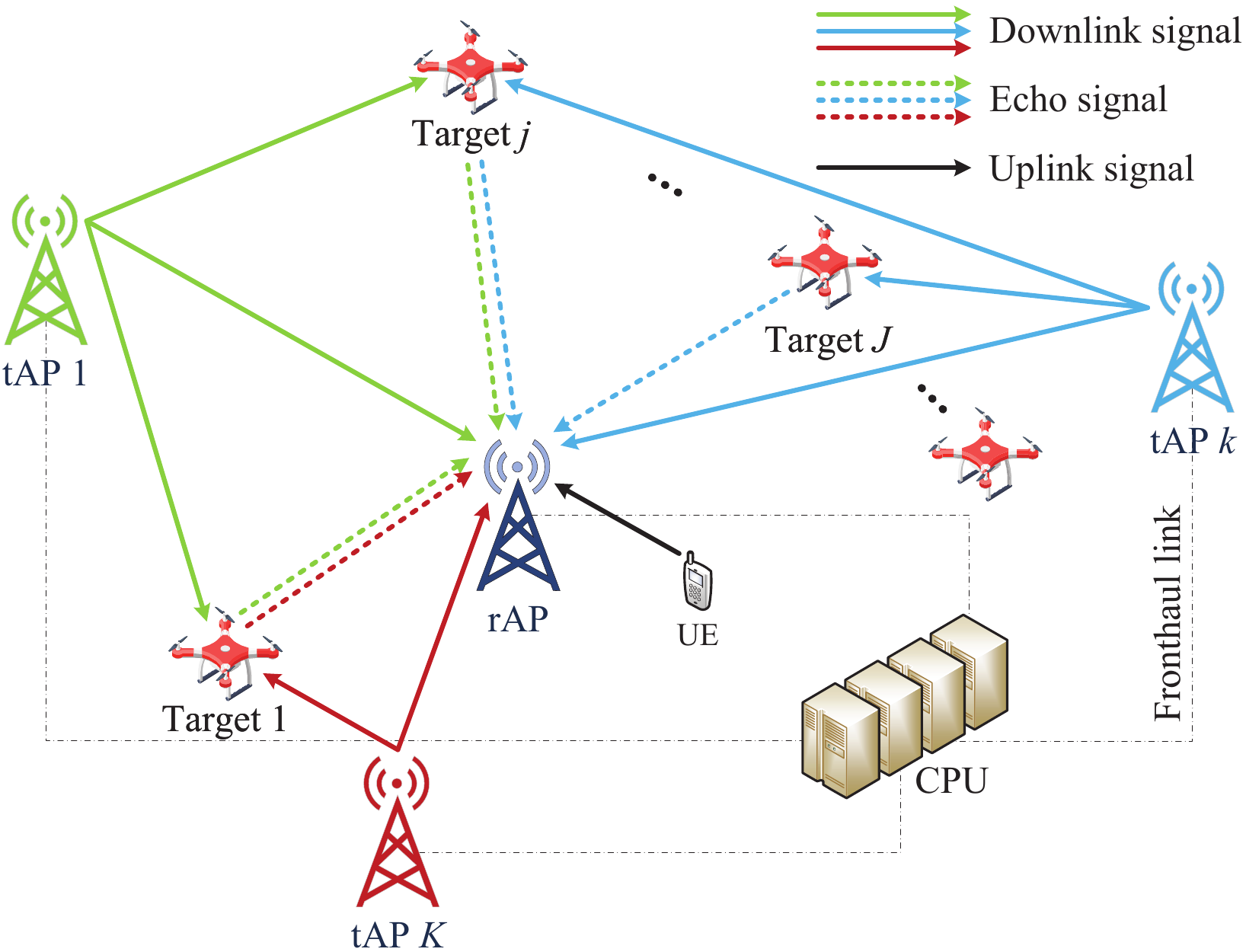}
	\caption{System model for a cell-free cooperative ISAC system with $K$ tAPs and $J$ targets.}
	\label{system_model}
\end{figure}

\begin{figure}\color{blue}
	\centering
	\includegraphics[width=0.9\linewidth]{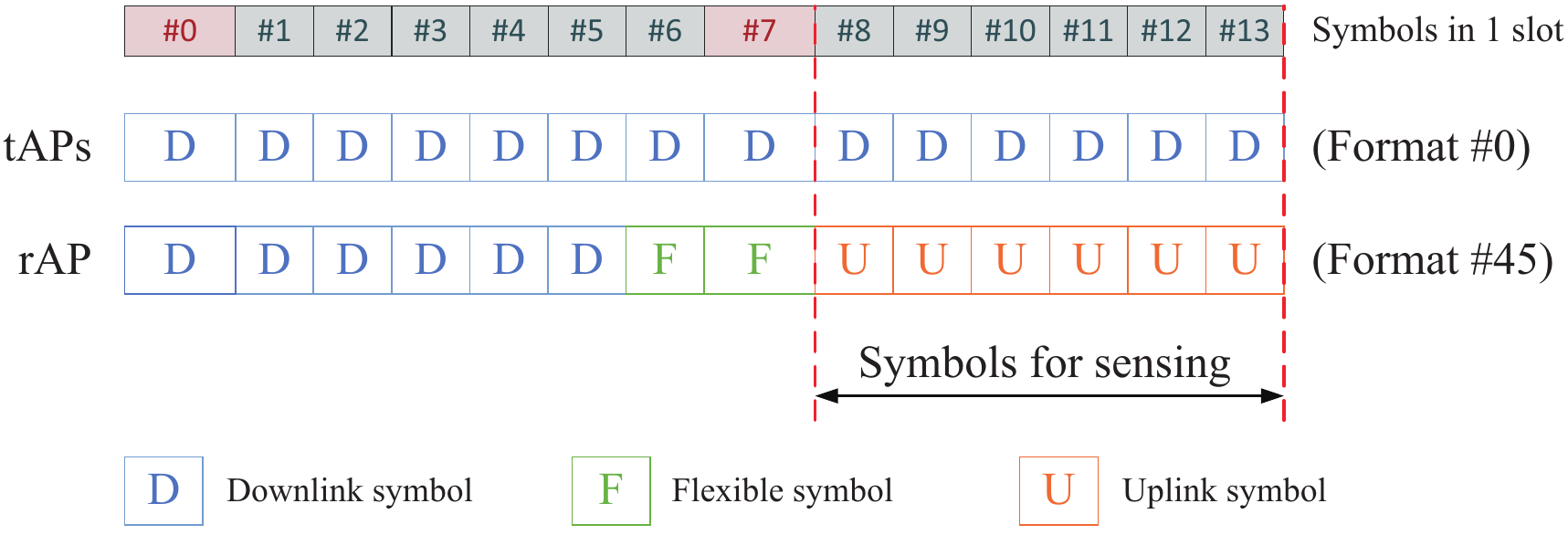}
	\caption{An example of feasible slot format configurations, where up to six symbols can be reused for sensing in each slot.}
	\label{SlotFormat_example}
\end{figure}

\subsection{Cell-Free Cooperative ISAC System}
{\color{blue}
Consider an OFDM-based cooperative ISAC system, which consists of multiple single-antenna APs as shown in Fig. \ref{system_model}.
Each AP operates on the time division duplexing (TDD) mode, which is the sole duplex mode employed for cellular networks at frequency bands exceeding 3 GHz as per 5G NR standards \cite{3GPP13810101,3GPP13810102}.
To mitigate the interference caused by the dense deployment, each AP exchanges signals with a UE on different bandwidth parts (BWPs), while the receive antenna of each AP covers all of the active BWPs.
Thus, an AP (called rAP) receiving uplink symbols can simultaneously receive and distinguish signals from those APs (called tAPs) that are transmitting downlink symbols.
All APs are connected to a CPU via fronthaul links, where the signals transmitted and received by all APs are centrally processed.
This configuration facilitates the sharing of comprehensive network information, such as the AP locations and the signal data.
In addition, the timing information can be sent to the APs via the fronthaul links to achieve synchronization among them.
}

{\color{blue}
Based on the flexible slot formats \cite{3GPP138213}, the uplink/downlink transmissions can be finely scheduled by the CPU.
In this context, we assume that in the duration of each sensing task, there is one active rAP along with $K$ active tAPs, and the selection of rAP is dynamically switched in response to the communication needs of UEs.
The sensing scheme for multi-rAP scenarios will be designed in Section \ref{subsubsec_MultirAP}.
Fig. \ref{SlotFormat_example} illustrates a configuration example to enable this transmission pattern, where tAPs and rAP utilize different frame structures. 
In this configuration, for the duration of the last six symbols of a time slot, each tAP transmits downlink symbols while the rAP receives uplink symbols from the UE and echo signals from targets.
Owing to the real-time sharing of the network information, the downlink information-bearing communication signals transmitted by the tAPs can be reused as sensing reference signals.
Similar to the bistatic radar, the time-delay information can be extracted from the echo signals received at the rAP.
Note that this ISAC framework does not require the APs to be equipped with additional sensing antennas to realize full-duplex transmission and reception of sensing signals \cite{9724258,10616023}, thereby enabling its implementation in the existing cell-free networks.
In this work, we focus on the device-free target localization based on the OFDM communication signals in the cooperative ISAC scenario.
}

\subsection{Signal Transmission Model}
{\color{blue}
It is assumed that the transmit power $P_{\mathrm t}$, the SCS $\Delta f^\mu$ and the number of active subcarriers $N_{\mathrm{c}}$ allocated for each AP are the same.
To facilitate signal processing, we only consider the cases where the reused OFDM symbols are continuous within each sensing task.
Suppose that $M$ OFDM symbols are reused for sensing.
Let us denote $s_{k,i,m}$ as the modulated symbol that modulates the $i$-th subcarrier of the $m$-th symbol, $i=0,1,\dots,N_{\mathrm{c}}-1$, $m=0,1,\dots,M-1$, transmitted by tAP $k$.
Then, the frequency-domain OFDM symbols transmitted from tAP $k$ can be denoted by 
\begin{align}
	\mathbf{S}_k=\left[ \begin{matrix}
		s_{k,0,0}&		s_{k,0,1}&		\cdots&		s_{k,0,M-1}\\
		s_{k,1,0}&		s_{k,1,1}&		\cdots&		s_{k,1,M-1}\\
		\vdots&		\vdots&		\ddots&		\vdots\\
		s_{k,N_{\mathrm{c}}-1,0}&		s_{k,N_{\mathrm{c}}-1,1}&		\cdots&		s_{k,N_{\mathrm{c}}-1,M-1}\\
	\end{matrix} \right] \in \mathbb{C} ^{N_{\mathrm{c}}\times M}.
\end{align}
Without loss of generality, we assume that the amplitude of $s_{k,i,m}$ satisfies $\left| s_{k,i,m} \right|=1$ for $\forall k,i,m$.
By operating zero padding and inverse fast Fourier transform (IFFT), the baseband time-domain OFDM signal transmitted from tAP $k$ is given by
\begin{align}
	x_{k}\left( t \right) =\sum_{m=0}^{M-1}{\sum_{i=0}^{N_{\mathrm{c}}-1}{s_{k,i,m}e^{j2\pi i\Delta f^{\mu}t}r\left( t-mT^{\mu} \right)}},
\end{align}
where $T^{\mu}$ is the OFDM symbol period including CP and $r\left( t \right)$ represents the pulse shaping filter.
Since the APs are typically deployed at open areas, there is a LoS-dominant multi-path sensing channel between each tAP and the rAP.

Let $\tau _{k,l}$ and $f_{\mathrm{D},k,l}$ denote the delay and Doppler frequency shift of the $l$-th path between tAP $k$ and the rAP, respectively, 
where the paths include the LoS path with index $l=0$ and the paths introduced by the targets with index $l=1,\dots,L$.
Denote by $J$ the number of targets to be sensed.
In this work, we assume that the echoes reflected by other scatterers are negligible, and thus, $L=J$.
In sensing applications within open environments, such as low-altitude drone detection, this assumption typically holds.
Then, the time-delay-domain sensing channel between tAP $k$ and the rAP is given by \cite{10616023}
\begin{align}\label{h_timedelay}
	h_k\left( t,\tau \right) =\sum_{l=0}^L{\alpha _{k,l}g^{\mathrm{Imp}} \left( \tau -\tau _{k,l}-\delta _{k}^{\tau} \right) e^{j2\pi \left( f_{\mathrm{D},k,l}+\delta _{k}^{f} \right) t}},
\end{align}
where $\alpha _{k,l}$ is the channel fading magnitude of the $l$-th path,
$g^{\mathrm{Imp}} \left( \tau \right)$ denotes the impulse function,
$\delta _{k}^{\tau}$ and $\delta _{k}^{f}$ are the unknown residual STO and CFO between tAP $k$ and the rAP after synchronization, respectively, which are assumed to be unchanged within the period of each sensing task \cite{10551673}.
By conducting the Fourier transform of delay $\tau$, we obtain the frequency-domain sensing channel on the $i$-th subcarrier as follows
\begin{align}\label{h_ki}
	h_{k,i}\left( t \right) =\sum_{l=0}^L{\alpha _{k,l}e^{-j2\pi i\Delta f^{\mu}\left( \tau _{k,l}+\delta _{k}^{\tau} \right)}e^{j2\pi \left( f_{\mathrm{D},k,l}+\delta _{k}^{f} \right) t}}.
\end{align}
Note that the received symbols are sampled at the rAP.
The discrete frequency-domain sensing channel for the $m$-th symbol can be expressed as
\begin{align}\label{h_kim}
	h_{k,i,m}=\sum_{l=0}^L{\alpha _{k,l}e^{-j2\pi i\Delta f^{\mu}\left( \tau _{k,l}+\delta _{k}^{\tau} \right)}e^{j2\pi mT^\mu\left( f_{\mathrm{D},k,l}+\delta _{k}^{f} \right)}}.
\end{align}
Then, the received frequency-domain OFDM symbols from tAP $k$ can be formulated as
\begin{align}
	\hat{s}_{k,i,m}=s_{k,i,m}h_{k,i,m}+z_{k,i,m},
\end{align}
where $\hat{s}_{k,i,m}$ is the received symbol on the $i$-th subcarrier of the $m$-th symbol, 
$z_{k,i,m}\sim \mathcal{C} \mathcal{N}\left( 0,\sigma _{z}^{2} \right)$ denotes the circularly symmetric complex Gaussian (CSCG) noise.
}

{\color{blue}
\section{Cooperative Sensing Scheme}\label{Sec_sens_scheme}
In this section, we apply the two-stage framework for the device-free localization of the passive targets. 
Specifically, the length of the scattered path related to each target is firstly estimated, and the results are used in Stage II for target localization.
}

\subsection{Stage I: Bistatic Range Measurement}
\begin{figure}\color{blue}
	\centering
	\includegraphics[width=0.88\linewidth]{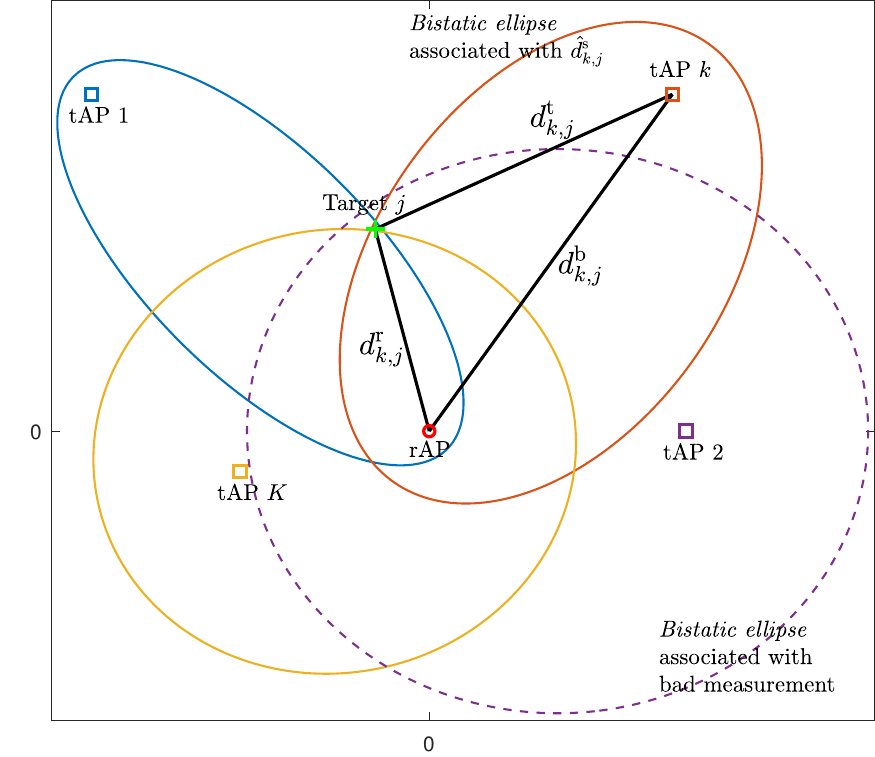}
	\caption{Illustration of bistatic ellipses corresponding to Target $j$, where an ill-conditioned measurement occurs.}
	\label{bistatic}
\end{figure}

{\color{blue}
In this stage, the \emph{bistatic ranges}, i.e., the ranges of the tAP-target-rAP paths, are estimated by extracting the time-delay information from channel $h_{k,i,m}$'s.
Without loss of generality, we assume that the rAP is located at $\left( 0,0 \right) $ as shown in Fig. \ref{bistatic}.
The coordinates of tAP $k$ and Target $j$ are denoted by $\mathbf{a}_k=\left[ a^x_k,a^y_k \right] ^{\mathrm{T}}$ and $\mathbf{q}_j=\left[ q^x_j,q^y_j \right] ^{\mathrm{T}}$, respectively, 
where the values of $\mathbf{a}_k$'s are known at the CPU.
Therefore, the distance between tAP $k$ and Target $j$ can be calculated as $d_{k,j}^{\mathrm{t}}=\left\| \mathbf{q}_j-\mathbf{a}_k \right\| _2$ and that between Target $j$ and the rAP is $d_{k,j}^{\mathrm{r}} = \left\| \mathbf{q}_j \right\| _2$.
Then, the bistatic range corresponding to Target $j$ and tAP $k$ is given by $d_{k,j}^{\mathrm{s}}=d_{k,j}^{\mathrm{t}}+d_{k,j}^{\mathrm{r}}$.
The range of the LoS path, also known as the \emph{baseline range}, between tAP $k$ and the rAP is given by $d_{k}^{\mathrm{b}}=\left\| \mathbf{a}_k \right\| _2$.
In addition, we define the \textit{bistatic ellipse} associated with the bistatic range $d_{k,j}^{\mathrm{s}}$ as the ellipse that satisfies: 
1) its foci are located at $\left( 0,0 \right)$ and $\mathbf{a}_{k}$, and 2) the equation $\left\| \mathbf{q} \right\| _2+\left\| \mathbf{q}-\mathbf{a}_k \right\| _2=d_{k,j}^{\mathrm{s}}$ holds for any point $\mathbf{q}$ on the ellipse.

\subsubsection{Sensing Information Extraction}
From \eqref{h_kim}, it can be noted that the LoS components of the channel $h_{k,i,m}$'s cannot be directly calculated based on the baseline ranges due to the unknown STO $\delta _{k}^{\tau}$'s and CFO $\delta _{k}^{f}$'s.
However, the delay and the Doppler frequency shift of the LoS path are known apriori as follows
\begin{gather}
	\tau _{k,0}=d_{k}^{\mathrm{b}}/c_0,
	\\
	f_{\mathrm{D},k,0} = 0,
\end{gather}
where $c_0$ denotes the speed of light.
Hence, given estimations of the LoS path delay $\hat\tau _{k,0}$ and Doppler frequency shift $\hat f_{\mathrm{D},k,0}$, the STO and CFO can be estimated as 
\begin{gather}
	\delta _{k}^{\tau}=\hat\tau _{k,0}-\tau _{k,0}=\hat\tau _{k,0}-d_{k}^{\mathrm{b}}/c_0,\label{STO}
	\\
	\delta _{k}^{f} = \hat f_{\mathrm{D},k,0}.\label{CFO}
\end{gather}
This motivates us to extract the delay and Doppler frequency shift of all $L+1$ paths between tAP $k$ and the rAP, and compensate for the STO and CFO by regarding the LoS path as a reference path.

Let us define a matrix of the received frequency-domain symbols from tAP $k$ as follows
\begin{align}
	\hat{\mathbf{S}}_{k}=\left[ \begin{matrix}
		\hat{s}_{k,0,0}&		\hat{s}_{k,0,1}&		\cdots&		\hat{s}_{k,0,M-1}\\
		\hat{s}_{k,1,0}&		\hat{s}_{k,1,1}&		\cdots&		\hat{s}_{k,1,M-1}\\
		\vdots&		\vdots&		\ddots&		\vdots\\
		\hat{s}_{k,N_{\mathrm{c}}-1,0}&		\hat{s}_{k,N_{\mathrm{c}}-1,1}&		\cdots&		\hat{s}_{k,N_{\mathrm{c}}-1,M-1}\\
	\end{matrix} \right] \in \mathbb{C} ^{N_{\mathrm{c}}\times M}.
\end{align}
Then, an estimation of the frequency-domain sensing channels can be obtained as $\hat{\mathbf{H}}_k=\mathbf{S}^{\ast}_k\odot \hat{\mathbf{S}}_k$, whose $\left(i+1,m+1\right)$-th element is given by
\begin{align}
	\left[ \hat{\mathbf{H}}_k \right] _{i+1,m+1}&=s_{k,i,m}^{\ast}s_{k,i,m}h_{k,i,m}+s_{k,i,m}^{\ast}z_{k,i,m}\notag 
	\\
	&=\sum_{l=0}^L{\alpha _{k,l}e^{-j2\pi i\Delta f^{\mu}\left( \tau _{k,l}+\delta _{k}^{\tau} \right)}
		e^{j2\pi mT^{\mu}\left( f_{\mathrm{D},k,l}+\delta _{k}^{f} \right)}}\notag
	\\
	&\quad+s_{k,i,m}^{\ast}z_{k,i,m}.
\end{align}
In the following, we extract the sensing information from $\hat{\mathbf{H}}_k$ based on the 2D-FFT method.
Denote by $\bar{\mathbf{H}}_k\in \mathbb{C} ^{N_{\mathrm{f}}\times N_{\mathrm{D}}}$ an extended version of $\hat{\mathbf{H}}_k$ with zero padding, where $ N_{\mathrm{D}}\geqslant M$ is the FFT size in the Doppler dimension.
By conducting IFFT along the subcarrier dimension and FFT along the symbol dimension on $\bar{\mathbf{H}}_k$, we can obtain a delay-Doppler spectrum $ \mathbf{P}_k $, whose $(p+1,q+1)$-th element is given by
\begin{align}\label{P_k}
	\left[ \mathbf{P}_k \right] _{p+1,q+1}=\frac{1}{N_{\mathrm{D}}}\sum_{m=0}^{N_{\mathrm{D}}-1}{\sum_{i=0}^{N_{\mathrm{f}}-1}{\left[ \bar{\mathbf{H}}_k \right] _{i+1,m+1}}} e^{jp\frac{2\pi}{N_{\mathrm{f}}}i}e^{-jq\frac{2\pi}{N_{\mathrm{D}}}m}.
\end{align}
The positions of the spectral peaks in $ \mathbf{P}_k $ indicate the delay and Doppler frequency shift of the paths between tAP $k$ and the rAP.
Nevertheless, as illustrated in Fig. \ref{Fig_2DFFT}, there exist several sidelobes around each spectral peak which may be higher than other peaks since $\alpha _{k,0}\gg\alpha _{k,1}\gg \alpha _{k,2}\gg\dots$ in most cases.
Therefore, we propose to extract the sensing information of each path and eliminate its corresponding spectral peak one-by-one. 
\begin{remark}
	\emph{Note that this work only focuses on the target localization.
		Hence, we only need to extract the delay information of the paths between the rAP and the tAPs, which can be achieved by conducting one-dimensional FFT on each column of $\bar{\mathbf{H}}_k$'s \cite{hanetalcellularnet}. 
		However, the proposed scheme, which simultaneously extracts delay and Doppler information from multiple symbols, offers the following advantages:
		(1)	Due to the varying speeds of targets, they are more distinguishable on a 2D delay-Doppler spectrum than on a one-dimensional delay spectrum, thereby reducing the probability of false detection;
		(2)	As it will be shown in the simulation results, when the number of reused symbols $M\gg 1$, the 2D-FFT method is able to correctly extract the delay information which can be longer than the CP duration.		
	}
\end{remark}

Define $\left(\tilde{p}_l,\tilde{q}_l\right)$ as the index of the spectral peak corresponding to the $l$-th path.
Note that the resolutions of the 2D-FFT method in the delay and Doppler frequency shift estimation are given by
\begin{gather}
	\Delta \tau =1/\left( N_{\mathrm{c}}\Delta f^{\mu} \right) ,\quad  \Delta f_{\mathrm{D}}=1/\left( MT^{\mu} \right) .
\end{gather}
To reduce the estimation error and mitigate the impact of ``error propagation" during the extraction-and-elimination process, we propose to solve the following problem:
\begin{subequations}\label{Prob_findpeak}
	\begin{alignat}{2}
		\max_{\check{p}_l ,\check{q}_l} \quad
		& \sum_{m=0}^{M-1}{\sum_{i=0}^{N_{\mathrm{c}}-1}{s_{k,i,m}e^{-j\check{p}_l\frac{2\pi i}{N_{\mathrm{f}}}}e^{j\check{q}_l\frac{2\pi m}{N_{\mathrm{D}}}}\cdot \hat{s}_{k,i,m}}}
		\\
		\mbox{s.t.}\quad
		&\tilde{p}_l-1\leqslant \check{p}_l\leqslant \tilde{p}_l+1, 
		\\
		&\tilde{q}_l-1\leqslant \check{q}_l\leqslant \tilde{q}_l+1.
	\end{alignat}
\end{subequations}
Problem \eqref{Prob_findpeak} can be solved by simultaneously performing binary search w.r.t. $\check{p}_l$ and $\check{q}_l$.
Based on the solutions $\check{p}_l$ and $\check{q}_l$ of Problem \eqref{Prob_findpeak}, we can obtain more accurate estimations of the channel fading magnitude $\alpha _{k,l}$, the delay  $\tau _{k,l}$ and the Doppler frequency shift $f_{\mathrm{D},k,l}$ of the $l$-th path as follows:
\begin{gather}
	\hat{\alpha}_{k,l}=\small{\frac{1}{N_{\mathrm{c}}M}}\sum_{m=0}^{M-1}{\sum_{i=0}^{N_{\mathrm{c}}-1}{s_{k,i,m}e^{-j\check{p}_l\frac{2\pi i}{N_{\mathrm{f}}}}e^{j\check{q}_l\frac{2\pi m}{N_{\mathrm{D}}}}\cdot \hat{s}_{k,i,m}}}, \label{tilde_alpha}
	\\
	\hat{\tau}_{k,l}=\check{p}_l/\left( N_{\mathrm{f}}\Delta f^{\mu} \right),
	\\
	\hat{f}_{\mathrm{D},k,l}=\check{q}_l/\left( N_{\mathrm{D}}T^{\mu} \right).
\end{gather}
Furthermore, the frequency-domain sensing channel component corresponding to the $l$-th path between tAP $k$ and the rAP can be estimated as
\begin{gather}\label{tildeH_component}
	\left[ \hat{\mathbf{H}}_{k,l}^{\prime} \right] _{i+1,m+1}=\hat{\alpha}_{k,l}e^{-j2\pi i\Delta f^{\mu}\left( \hat{\tau}_{k,l}+\delta _{k}^{\tau} \right)}e^{j2\pi mT^{\mu}\left( \hat{f}_{\mathrm{D},k,l}+\delta _{k}^{f} \right)}.
\end{gather}
\begin{algorithm}[tbh]\color{blue}
	\caption{Sensing Information Extraction Algorithm}
	\label{alg_2DFFT}
	\begin{algorithmic}[1] 
		\Require 
		$K$, $L$, $\lbrace \hat{\mathbf{H}}_k\rbrace _{k=1}^K$ and $\lbrace d_{k}^{\mathrm{b}}\rbrace\rbrace _{k=1}^K $.
		\Ensure 
		Estimations of delay $\hat{\tau}_{k,l}$'s, Doppler frequency shift $\hat{f}_{\mathrm{D},k,l}$'s, STO $\delta _{k}^{\tau}$'s and CFO $\delta _{k}^{f}$'s.
		\ForAll{$k$}
		\State
		$l\gets 0$;
		\While{$l\leqslant L$}
		\State
		Calculate $ \mathbf{P}_k $ via \eqref{P_k};
		\State
		Obtain $\left(\tilde{p}_l,\tilde{q}_l\right)=\max _{\left( p,q \right)}\left\{ \left[ \mathbf{P}_k \right] _{p+1,q+1} \right\}$;
		\State
		Obtain $\check{p}_l$ and $\check{q}_l$ by solving Problem \eqref{Prob_findpeak};
		\State
		Calculate $\hat\alpha _{k,l}$, $\hat\tau _{k,l}$, $\hat f_{\mathrm{D},k,l}$ and $\hat{\mathbf{H}}_{k,l}^{\prime}$ via \eqref{tilde_alpha}-\eqref{tildeH_component};
		\State
		$\hat{\mathbf{H}}_k\gets\hat{\mathbf{H}}_k-\hat{\mathbf{H}}_{k,l}^{\prime}$;
		\State 
		$l\gets l+1$;
		\EndWhile
		\State
		Calculate $\delta _{k}^{\tau}$ and $\delta _{k}^{f}$ via \eqref{STO} and \eqref{CFO};
		\EndFor 
	\end{algorithmic}
\end{algorithm}

\subsubsection{Algorithm Development}
Based on the above discussions, we summarize the proposed method for estimating sensing information in Algorithm \ref{alg_2DFFT}.
The computational complexity of Algorithm \ref{alg_2DFFT} is analyzed as follows:
The complexity of calculating $ \mathbf{P}_k $ by 2D-FFT is of order $\mathcal{O} \left( N_{\mathrm{D}}N_f\log \left( N_{\mathrm{D}}N_f \right) \right) $.
In addition, the complexity of solving Problem \eqref{Prob_findpeak} is of order $\mathcal{O} \left( r^{\mathrm{BS}}_{\mathrm{iter}}N_{\mathrm{c}}M \right) $, where $r^{\mathrm{BS}}_{\mathrm{iter}}$ denotes the number of iterations required by the binary search algorithm.
Therefore, the total complexity of Algorithm \ref{alg_2DFFT} can be expressed as $\mathcal{O} \left( KLN_{\mathrm{D}}N_f\left( \log \left( N_{\mathrm{D}}N_f \right) +r^{\mathrm{BS}}_{\mathrm{iter}} \right) \right)$.

Given the estimation $\hat{\tau}_{k,l}$'s, the CPU can obtain a set of bistatic range measurements corresponding to each tAP as follows
\begin{align}\label{D_k}
	\!\mathcal{D} _k=\left\{ \hat{d}_{k,j}^{\mathrm{s}}\left| \hat{d}_{k,j}^{\mathrm{s}}=c_0\left( \hat{\tau}_{k,j}-\delta _{k}^{\tau} \right) , j= \right. 1,\dots ,J \right\} ,\;\forall k.
\end{align}
In the next subsection, we propose a method to estimate the locations of the targets based on the measurements $\left\lbrace \mathcal{D} _k\right\rbrace _{k=1}^K$.

\begin{figure}[t]\color{blue}
	\centering  
	\subfigbottomskip=2pt 
	\subfigcapskip=-5pt 
	\subfigure[LoS path extraction]{
		\includegraphics[width=0.45\linewidth]{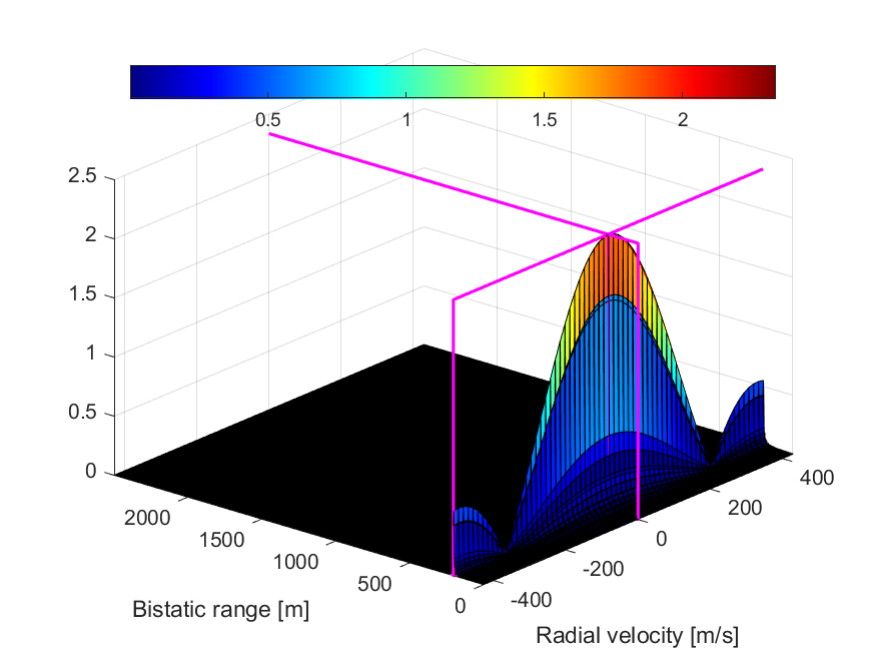}}
	\subfigure[1st scattered path extraction]{
		\includegraphics[width=0.45\linewidth]{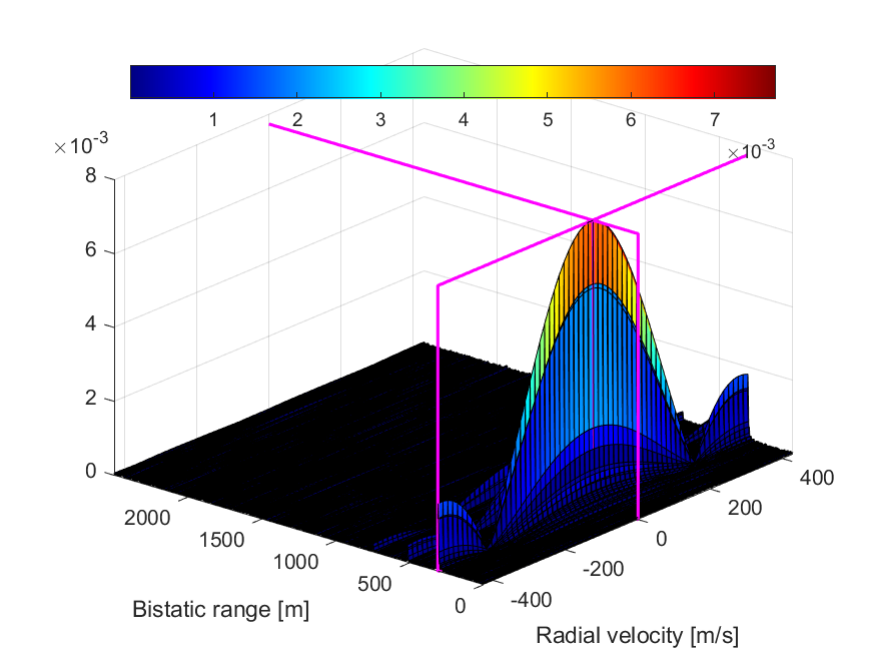}}
	\\
	\subfigure[2nd scattered path extraction]{
		\includegraphics[width=0.45\linewidth]{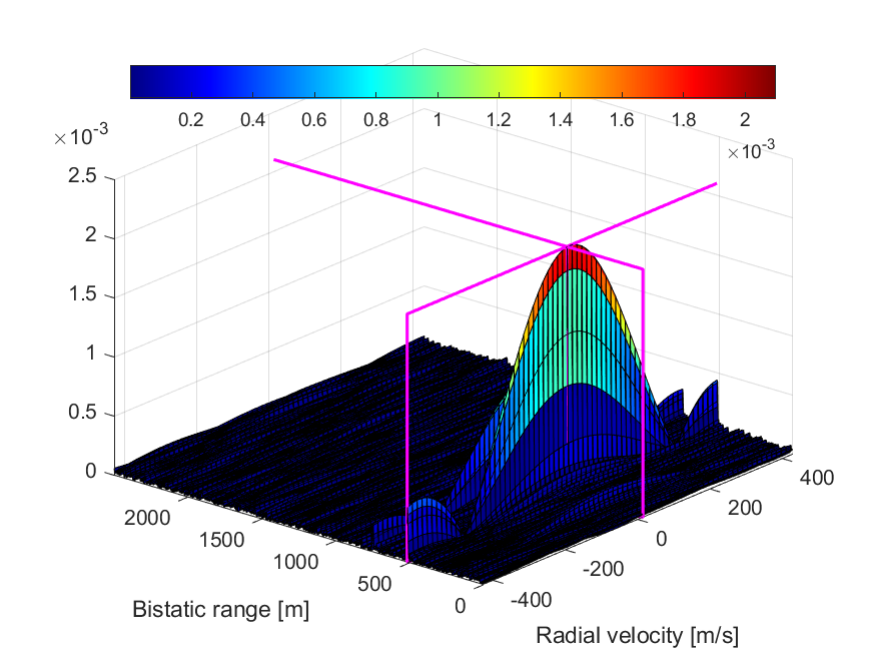}}
	\subfigure[3rd scattered path extraction]{
		\includegraphics[width=0.45\linewidth]{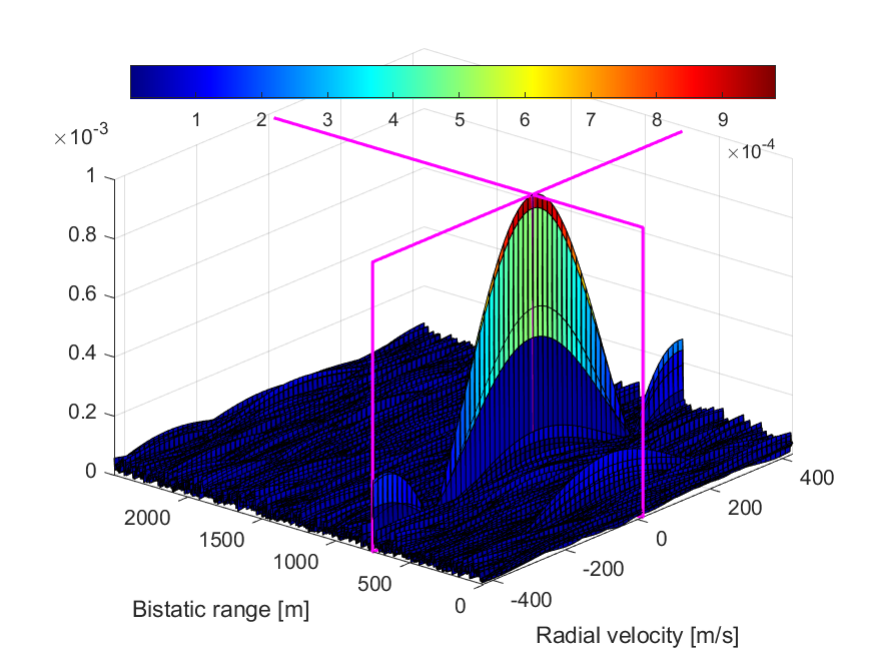}}
	\caption{An example for the extraction-and-elimination process of each path when the number of targets is $J=3$.}
	\label{Fig_2DFFT}
\end{figure}

\subsubsection{Performance Evaluation}\label{evaluateAlg1}
To evaluate the performance of the proposed algorithm in the sensing information extraction and the compensation for STO and CFO, we consider a scenario with a pair of tAP and rAP located at $(-100,0)$ and $(100,0)$, respectively.
The SCS $\Delta f^\mu$, transmit power $P_{\mathrm{t}}$ and carrier frequency $f_{\mathrm{c}}$ of the tAP are set to 30 KHz, 45 dBm and 4.9 GHz, respectively.
Each target with a radar cross-section (RCS) of 1 m$^2$ is uniformly and randomly generated within a circular area centred at $(0,0)$ with a radius of 400 m.
The channel fading magnitude of each path is generated following the free-space attenuation given in \cite{ITURRECP525}.
Fig. \ref{Fig_2DFFT} shows an example for the process of path extraction and elimination of Algorithm \ref{alg_2DFFT} when the number of targets is $J=3$.
It can be seen that the proposed algorithm achieves high-accuracy delay and Doppler frequency shift estimation, and the sidelobes can be effectively eliminated.
}

\begin{figure}[t]\color{blue}
	\centering
	\includegraphics[width=0.9\linewidth]{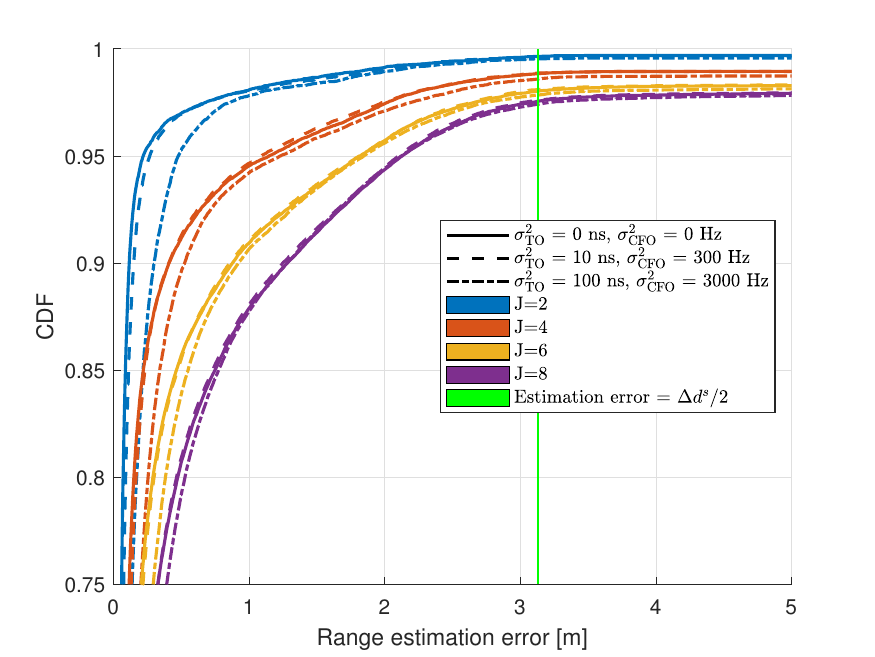}
	\caption{Accuracy of the bistatic range estimation.}
	\label{Fig_RangeEstiAccu}
\end{figure}

{\color{blue}
Then, we consider three levels of synchronization quality between the APs, indicated by the variance of STO $\sigma _{\mathrm{STO}}^{2}$ in nanosecond (ns) and the variance of CFO $\sigma _{\mathrm{CFO}}^{2}$ in Hz.
Fig. \ref{Fig_RangeEstiAccu} shows the cumulative distribution function (CDF) of the bistatic range estimation error with $J=[2,4,6,8]$.
Note that the attenuation of the LoS path is significantly lower than that of other paths.
When the length of a scattered path is close to that of the LoS path, its corresponding spectral peak is likely to be obscured by the peak corresponding to the LoS path.
It is challenging to correctly extract such scattered path.
Therefore, we define the \textit{sensing blind zone} (SBZ) of a tAP-rAP pair as the small elliptical region where any target located inside satisfies $d_{k,j}^{\mathrm{s}}\leqslant d_{k}^{\mathrm{b}}+3.5\Delta d^{\mathrm{s}}$.\footnote{\color{blue}
	In this work, the parameter is set to 3.5 because our simulations indicate that when the bistatic range of a target satisfies this inequality, the corresponding spectral peak is likely to be obscured, leading to estimation failure.}
Here, $\Delta d^{\mathrm{s}}$ is the \textit{bistatic range resolution} of the 2D-FFT algorithm given as follows
\begin{align}
	\Delta d^{\mathrm{s}}=c_0/\left( N_{\mathrm{c}}\Delta f^{\mu} \right).
\end{align}
Fig. \ref{Fig_RangeEstiAccu} only accounts for targets that are not within this SBZ.
It can be firstly observed that the estimation error experiences a slight increase as the number of targets increases.
This is because, when there are many targets, the residual effect from path elimination becomes greater, and it is likely that some targets have similar bistatic range.
In addition, we can see that the accuracy of range estimation is nearly unaffected by STO and CFO, which confirms the effectiveness of estimating STO and CFO by leveraging the LoS path.
}

\subsection{Stage II: Target Location Estimation}
In Stage II, the CPU jointly estimates the locations of the targets by utilizing the bistatic range measurements $\left\lbrace \mathcal{D} _k\right\rbrace _{k=1}^K$ obtained in Stage I.
As illustrated in Fig. \ref{bistatic}, the coordinates of a target can only be determined when multiple bistatic ellipses intersect at one point.
{\color{blue}
However, due to the unknown measurement error of bistatic ranges as shown in Fig. \ref{Fig_RangeEstiAccu}, a common intersection of the ellipses is unlikely to exist.
In addition, based on the discussion in the previous subsection, \textit{ill-conditioned measurements} with significant errors may occur if:
(1) a target is located within the SBZ of a tAP-rAP pair; or
(2) two targets are approximately located on the same bistatic ellipse.\footnote{\color{blue}
	Under condition (2) with limited $M$, the spectral peaks corresponding to the two targets are likely to merge, leading to Algorithm \ref{alg_2DFFT} eliminating two scattered path components from $\hat{\mathbf{H}}_k$ at once but generating only one delay estimation $\hat\tau _{k,l}$.
	In such cases, when Algorithm \ref{alg_2DFFT} attempts to extract $\hat\tau _{k,L}$, there is no scattered path component left in matrix $\hat{\mathbf{H}}_k$.}
Therefore, an ML-based algorithm is proposed for ill-conditioned measurement elimination and target localization in this subsection.

\subsubsection{Problem Formulation}
Firstly, we introduce a measurement error model as follows to approximate the correct bistatic range measurements \cite{9724258}:
\begin{align}\label{hat_d_kj_approx}
	\hat{d}_{k,j}^{\mathrm{s}}=d_{k,j}^{\mathrm{s}}+\varepsilon _{k,j},\; j\in\mathcal{J} \backslash {\mathcal{B}}_k,
\end{align}
where $\varepsilon _{k,j}\sim \mathcal{C} \mathcal{N} \left( 0,\sigma _{k,j}^{2} \right)$  is the error for estimating $d_{k,j}^{\mathrm{s}}$, 
$\mathcal{J} =\left\{ 1,\dots ,J \right\} $, and 
the set $\mathcal{B} _k$ is defined as the unknown index set of ill-conditioned measurements in $\mathcal{D} _k$.
Let $\mathcal{D} _k\left(t \right)$ denote the $t$-th largest element in set $\mathcal{D} _k$. 
In particular, we define $\mathcal{D} _k\left( 0 \right)=\oslash $.
Then, the matching between Target $j$ and the range measurement $\mathcal{D} _k\left( f_{k,j} \right)$ can be represented as a data association variable $f_{k,j}\in\mathcal{J} \cup \left\{ 0 \right\}$.
Note that the correct matchings $\mathcal{F} ^*=\left\{ f_{k,j}^{*}\left| \forall k,j \right. \right\} $ satisfying $\hat{d}_{k,j}^{\mathrm{s}}=\mathcal{D} _k\left( f^*_{k,j} \right) $ for $\forall j\in \mathcal{J} \backslash {\mathcal{B}}_k$ is not known at the CPU.
}
However, based on the measurement error model in \eqref{hat_d_kj_approx}, the probability distribution function of the event $\hat{d}_{k,j}^{\mathrm{s}}=\mathcal{D} _k\left( f_{k,j} \right) $, given $\mathbf{q}_j $ and $f_{k,j}$, can be formulated as 
\begin{align}\label{pdf}
	&p\left( \mathcal{D} _k\left( f_{k,j} \right) \left| \mathbf{q}_j,f_{k,j} \right. \right) \notag 
	\\
	&=\frac{1}{\sqrt{2\pi}\sigma _{k,j}}e^{-\frac{\left( \mathcal{D} _k\left( f_{k,j} \right) -\left\| \mathbf{q}_j \right\| _2-\left\| \mathbf{q}_j-\mathbf{a}_k \right\| _2 \right) ^2}{2\sigma _{k,j}^{2}}},\;
	\forall k,\, j\in\mathcal{J} \backslash {\mathcal{B}}_k.
\end{align}

Define $\mathcal{Q} =\left\{ \mathbf{q}_j\left| \forall j \right. \right\} $ and  $\mathcal{F} =\left\{ f_{k,j}\left| \forall k, j \right. \right\} $.
Based on \eqref{pdf}, the ML problem can be formulated as
{\color{blue}
\begin{subequations}\label{Prob_origin}
	\begin{alignat}{2}
		\min_{\mathcal{Q} ,\mathcal{F}} \;
		& \sum_{j\in \mathcal{J}}{\sum_{k\in \mathcal{K}_{j}}{\frac{\left( \mathcal{D} _k\left( f_{k,j} \right) -\left\| \mathbf{q}_j \right\| _2-\left\| \mathbf{q}_j-\mathbf{a}_k \right\| _2 \right) ^2}{2\sigma _{k,j}^{2}}}}
		\\
		\mbox{s.t.}\;\,
		&\left\{ f_{k,j}\left| \forall j\in \mathcal{J} \backslash {\mathcal{B}} _k \right. \right\} =\mathcal{J} \backslash {\mathcal{B}} _k,\,\forall k,\label{constrain_f_range}
		\\
		&f_{k,j}=0,\, \forall k, \, j\in {\mathcal{B}} _k,
	\end{alignat}
\end{subequations}
where $\mathcal{K}_{j}=\left\{ k\left| f_{k,j}>0 \right. \right\} $.
The constraint \eqref{constrain_f_range} ensures that different correct measurements in $\mathcal{D} _k$ are associated with different targets. 

Next, we consider the ideal case of estimation with $\mathcal{B} _k=\oslash$.}
Note that Problem \eqref{Prob_origin} is non-convex and difficult to solve.
However, given a feasible data association solution $\tilde{\mathcal{F}}=\left\{ \tilde f_{k,j}\left| \forall k, j \right. \right\} $ satisfying constraint \eqref{constrain_f_range}, it can be transformed as
\begin{align}\label{Prob_givenF}
	\min_{\mathcal{Q}} \;
		\sum_{j\in \mathcal{J}}{\sum_{k\in\tilde{\mathcal{K}}_{j}}{\frac{\left( \mathcal{D} _k\left( {\tilde{f}_{k,j}} \right) -\left\| \mathbf{q}_j \right\| _2-\left\| \mathbf{q}_j-\mathbf{a}_k \right\| _2 \right) ^2}{2\sigma _{k,j}^{2}}}},
\end{align}
where $\tilde{\mathcal{K}}_{j}=\left\{ k\left| \tilde f_{k,j}>0 \right. \right\} $.
Then, the optimal solution to Problem \eqref{Prob_givenF} can be obtained by separately solving $J$ subproblems, where the $j$-th subproblem is given by
\begin{align}\label{Prob_sub}
		\min_{\mathbf{q}_j } \;
		 \sum_{k\in\tilde{\mathcal{K}}_{j}}{\frac{\left( \mathcal{D} _k\left( {\tilde{f}_{k,j}} \right) -\left\| \mathbf{q}_j \right\| _2-\left\| \mathbf{q}_j-\mathbf{a}_k \right\| _2 \right) ^2}{2\sigma _{k,j}^{2}}}.
\end{align}
Subproblem \eqref{Prob_sub} is a non-linear least squared problem, which can be efficiently solved by the Gauss-Newton algorithm \cite{256302}.
Additionally, a closed-form solution for Problem \eqref{Prob_sub} can be obtained by the spherical-intersection (SX) algorithm \cite{6129656}, providing a good initial value to ensure the convergence of the Gauss-Newton algorithm.
{\color{blue}
Nonetheless, the optimal data association solution $\mathcal{F}^*$ cannot be found by the existing method except the exhaustive search method, whose complexity is prohibitively high since there are $\prod_{k=2}^K{J!}$ feasible data association solutions, each of which corresponds to $J$ subproblems as shown in \eqref{Prob_sub}.

Recall that ill-conditioned measurements may occur in practical scenarios.
In the case with $\mathcal{B} _k\ne\oslash$, the target location estimations obtained by solving Problem \eqref{Prob_givenF} may be subject to considerable errors.
Note that the number of ill-conditioned measurements $\left| \mathcal{B} _k \right|$ is unknown.
Hence, the values of ${\mathcal{B}} _k$'s cannot be determined by the exhaustive search method.
To address the aforementioned issues, in the following, we propose a low-complexity algorithm for eliminating the ill-conditioned measurements and solving Problem \eqref{Prob_origin}.
}

\subsubsection{Rough Estimation}

\begin{figure}
	\centering
	\includegraphics[width=0.85\linewidth]{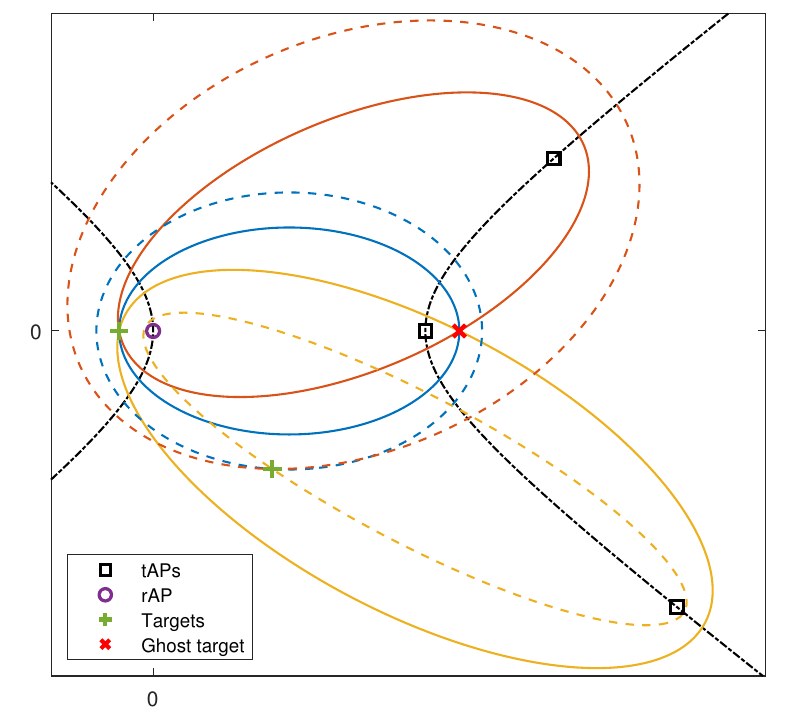}
	\caption{An example with ghost target.}
	\label{Fig_ghost}
\end{figure}

In our considered device-free sensing scheme, all the target locations are estimated by using the bistatic ranges extracted from the reflected communication signals.
In this case, the deployment locations of the APs may lead to estimation errors, even though the bistatic range measurements are precise.
As illustrated in Fig \ref{Fig_ghost}, when the bistatic ellipses (marked with solid lines) have two common intersection points, a \textit{ghost target} may be detected instead of the real one.
In the following, we firstly introduce a lemma to show the sufficient conditions when there are no ghost targets in the bistatic location estimation:
\vspace{0.1 cm}
\begin{lemma}\label{lemma_rough_estimaion}
	\emph{When there is no error in the measurement of bistatic ranges, a sufficient condition for ensuring a unique common intersection among multiple bistatic ellipses is that there does not exist a hyperbola where all tAPs lie on one branch while the rAP is located on the other.}
	
\noindent\textit{Proof:} 
We prove Lemma \ref{lemma_rough_estimaion} by demonstrating its contrapositive, i.e., ``When multiple bistatic ellipses have two or more common intersections, there must exist a hyperbola such that all tAPs lie on one branch and the rAP is located on the other."
Recall that the rAP is located at $\left( 0,0 \right) $.
Assuming that $\mathbf{r}_1$ and $\mathbf{r}_2$ are two of the common intersections, we have
\begin{align}\label{eq_ellipse}
	\left\| \mathbf{r}_1-\mathbf{a}_k \right\| _2+\left\| \mathbf{r}_1 \right\| _2=\left\| \mathbf{r}_2-\mathbf{a}_k \right\| _2+\left\| \mathbf{r}_2 \right\| _2,\; \forall k.
\end{align}
Then, we have
\begin{align}\label{eq_ellipse2}
	\left\| \mathbf{r}_1-\mathbf{a}_k \right\| _2-\left\| \mathbf{r}_2-\mathbf{a}_k \right\| _2=-\left( \left\| \mathbf{r}_1 \right\| _2-\left\| \mathbf{r}_2 \right\| _2 \right) ,\; \forall k.
\end{align}
The equations in \eqref{eq_ellipse2} show that all tAPs are located on a branch of a hyperbola with $\mathbf{r}_1$ and $\mathbf{r}_1$ as foci, while the rAP is on the other branch.
Hence, we arrive at the contrapositive of Lemma \ref{lemma_rough_estimaion}, and thus the proof is completed.
$\hfill\blacksquare$
\end{lemma}

\begin{figure}
	\centering
	\includegraphics[width=0.9\linewidth]{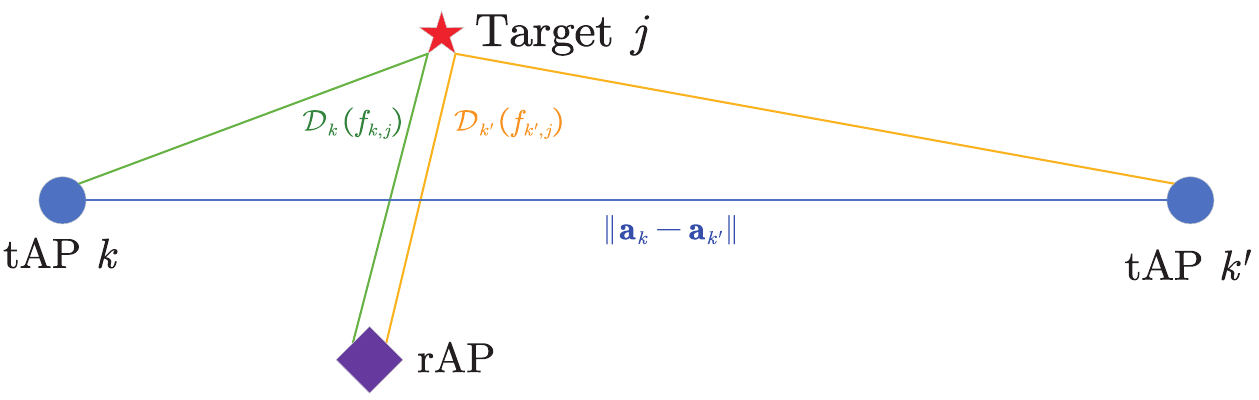}
	\caption{Illustration of triangle inequalities.}
	\label{Fig_TriangleInequatity}
\end{figure}

Lemma \ref{lemma_rough_estimaion} provides engineering guidance for the deployment of APs in cooperative networks to avoid the appearance of ghost targets.
In this work, we focus on the scenario that the APs are appropriately deployed.
In this case, the condition of the unique intersection given in Lemma \ref{lemma_rough_estimaion} is assumed to be satisfied, which means only three bistatic range measurements are required for a rough location estimation.
{\color{blue}
To estimate the target position while eliminating ill-conditioned measurements, we propose to associate the range measurements with the targets based on the greedy algorithm.
Firstly, we estimate the locations of $\tilde{J}$ targets based on three measurement sets, denoted by $\mathcal{D} _{c_1}$, $\mathcal{D} _{c_2}$ and $\mathcal{D} _{c_3}$, where ${c_1}$, ${c_2}$ and ${c_3}$ are different elements in $\left\lbrace 1,2,\dots,K\right\rbrace $ and $\tilde{J}$ is the number of targets with no ill-conditioned measurements in the three sets.
Given the SBZs of the APs and the distribution range of targets, we can readily eliminate a small number of ill-conditioned measurements which are extremely large or small.
Hence, we have $ 0\leqslant\tilde{J}\leqslant\min_{ k\in \mathcal{C}} \left\{ \left| \mathcal{D} _k \right| \right\} $.
A method for determining the value of $\tilde{J}$ will be discussed in Remark \ref{remark_TildeJ}.
Let us define a combination $ \mathcal{C}=\left\lbrace c_1,c_2,c_3\right\rbrace $.
Similar to Problem \eqref{Prob_origin}, the rough estimation problem for locating the $\tilde{J}$ targets can be formulated as
\begin{subequations}\label{Prob_origin3}
	\begin{alignat}{2}
		\min_{\mathcal{Q} _{\tilde{J}}^{\left( 3 \right)},\mathcal{F} _{\tilde{J}}^{\left( 3 \right)}}\;
		& \sum_{j=1}^{\tilde{J}}{\sum_{k\in \mathcal{C}}{\frac{\left( \mathcal{D} _k\left( f_{k,j} \right) -\left\| \mathbf{q}_{j}^{\left( 3 \right)} \right\| _2-\left\| \mathbf{q}_{j}^{\left( 3 \right)}-\mathbf{a}_k \right\| _2 \right) ^2}{2\sigma _{k,j}^{2}}}}
		\\
		\mbox{s.t.}\;\,
		&f_{k,j}\in \mathcal{J} ,\; k\in \mathcal{C} ,\,j\leqslant \tilde{J},\label{constrain_f3_range}
		\\
		&f_{k,j} \ne f_{k,j\prime}\;\mbox{if}\; j\ne j\prime,\,  k\in \mathcal{C}, j\leqslant \tilde{J},\, j\prime\leqslant \tilde{J}\label{constrain_f3_notrepeat}
	\end{alignat}
\end{subequations}
where $\mathcal{Q}_{\tilde{J}} ^{\left( 3 \right)}=\left\{ \mathbf{q}_{j}^{\left( 3 \right)}\left| j\leqslant \tilde{J} \right. \right\} $ and $\mathcal{F} _{\tilde{J}}^{\left( 3 \right)}=\left\{ f_{k,j}\left| k\in \mathcal{C},j\leqslant \tilde{J} \right. \right\} $.
}
Given a feasible data association solution  $\tilde{\mathcal{F}}_{\tilde{J}}^{\left( 3 \right)}=\left\{ \tilde{f}_{k,j}\left| k\in \mathcal{C},j\leqslant \tilde{J} \right. \right\} $ that satisfies constraints \eqref{constrain_f3_range} and \eqref{constrain_f3_notrepeat}, Problem \eqref{Prob_origin3} can be transformed to
\begin{align}\label{Prob_sub3}
	\!\!\!\min_{\mathcal{Q}_{\tilde{J}}^{\left( 3 \right)}} \;
	\sum_{j=1}^{\tilde{J}}{\sum_{k\in \mathcal{C}}{\frac{\left( \mathcal{D} _k\left( \tilde{f}_{k,j} \right) -\left\| \mathbf{q}_{j}^{\left( 3 \right)} \right\| _2-\left\| \mathbf{q}_{j}^{\left( 3 \right)}-\mathbf{a}_k \right\| _2 \right) ^2}{2\sigma _{k,j}^{2}}}}.\!
\end{align}
Problem \eqref{Prob_sub3} can also be decoupled into ${\tilde{J}}$ subproblems and solved by the SX and Gauss-Newton algorithms.
Define the objective function of Problem \eqref{Prob_sub3} as $G( \tilde{\mathcal{F}}_{\tilde{J}}^{\left( 3 \right)},\mathcal{Q}_{\tilde{J}} ^{\left( 3 \right)} ) $.
After searching all the feasible $\tilde{\mathcal{F}}_{\tilde{J}}^{\left( 3 \right)}$'s,
we select the solution that minimizes $G( \tilde{\mathcal{F}}_{\tilde{J}}^{\left( 3 \right)},\mathcal{Q}_{\tilde{J}} ^{\left( 3 \right)} ) $, denoted by $\left\{ \tilde{\mathcal{F}}_{\tilde{J}}^{\left( 3 \right) ^*},\mathcal{Q} ^{\left( 3 \right) ^*} \right\} $, as the rough estimation of the $\tilde{J}$ targets.
{\color{blue}
	Note that we do not need to solve Problem \eqref{Prob_sub3} for each feasible $\tilde{\mathcal{F}}_{\tilde{J}}^{\left( 3 \right)}$.
	Specifically, as illustrated in Fig. \ref{Fig_TriangleInequatity}, a data association solution is incorrect if it dose not satisfy the following triangle inequalities
	\begin{align}\label{triangle_inequal}
		&\left| \mathcal{D} _k\left( f_{k,j} \right) -\mathcal{D} _{k\prime}\left( f_{k\prime,j} \right) \right|<\left\| \mathbf{a}_k-\mathbf{a}_{k\prime} \right\|,\notag
		\\
		& k,k\prime\in \mathcal{C} ,\, j\leqslant \tilde{J}.
	\end{align}
	Therefore, a large proportion of the feasible solutions can be readily eliminated.
	}

\begin{remark}\label{remark_TildeJ}
	\emph{Let us define the estimated bistatic range corresponding to tAP $k$ and its location estimation $\mathbf{q}_{j}$ as follows
	\begin{align}\label{ds_esti}
		\bar{d}_{k}^{\mathrm{s}}\left( \mathbf{q}_{j}\right) = \left\| \mathbf{q}_{j} \right\| _2+\left\| \mathbf{q}_{j}-\mathbf{a}_k \right\| _2.
	\end{align}
	To determine the true value of $\tilde{J}$, we propose to initialize it as $\min_{ k\in \mathcal{C}} \left\{ \left| \mathcal{D} _k \right| \right\} $.
	After $\left\lbrace \tilde{\mathcal{F}}_{\tilde{J}}^{\left( 3 \right) ^*},\mathcal{Q} ^{\left( 3 \right) ^*}\right\rbrace $ is obtained, we can calculate the following indicator
	\begin{align}\label{indicator}
		&\zeta \left( \tilde{\mathcal{F}}_{\tilde{J}}^{\left( 3 \right) ^*},\mathcal{Q} ^{\left( 3 \right) ^*} \right) =\notag
		\\
		&\max _{\tilde{f}_{k,j}^{*}\in \tilde{\mathcal{F}}_{\tilde{J}}^{\left( 3 \right) ^*},\mathbf{q}_{j}^{\left( 3 \right) ^*}\in \mathcal{Q} ^{\left( 3 \right) ^*}}\left\lbrace \left| \mathcal{D} _k\left( \tilde{f}_{k,j}^{*} \right) -\bar{d}_{k}^{\mathrm{s}}\left( \mathbf{q}_{j}^{\left( 3 \right) ^*} \right) \right|\right\rbrace .
	\end{align}
	{\color{blue}
	Given a threshold $\zeta ^{\mathrm{th}}$, there exist ill-conditioned measurements in $\left\{ \mathcal{D} _k \right\} _{k\in \mathcal{C}}$ for the $\tilde{J}$ targets if $\zeta \left( \tilde{\mathcal{F}}_{\tilde{J}}^{\left( 3 \right) ^*},\mathcal{Q} ^{\left( 3 \right) ^*} \right)>\zeta ^{\mathrm{th}}$.
	Then, we can update the value of $\tilde{J}$ as $\tilde{J}\gets\tilde{J}-1$ and repeat the above procedure until $\tilde{J}=0$ or $\zeta \left( \tilde{\mathcal{F}}_{\tilde{J}}^{\left( 3 \right) ^*},\mathcal{Q} ^{\left( 3 \right) ^*} \right)\leqslant\zeta ^{\mathrm{th}}$.}}
\end{remark}

\subsubsection{Accurate Estimation}
Based on the rough estimation $\mathcal{Q} ^{\left( 3 \right) ^*}$, we can associate the $\tilde{J}$ targets with the measured bistatic range $\hat{d}_{k,j}^{\mathrm{s}}$'s corresponding to other tAPs for accurate location estimation.
Specifically, the data association solutions corresponding to $\left\{ \mathcal{D} _k \right\} _{k\notin \mathcal{C}}$, denoted by $\tilde{\mathcal{F}} _{\tilde{J}}^{\mathrm{rest}^*}= \left\{ \tilde{f}_{k,j}^{*}\left| k\notin \mathcal{C},j\leqslant \tilde{J} \right. \right\} $, are set as follows
\begin{align}\label{fkj_others}
	&\tilde{f}^*_{k,j}=\begin{cases}
		0,\quad  \mathrm{if}\underset{\hat{d}_{k,j}^{\mathrm{s}}\in \mathcal{D} _k}{\text{min}}\left\{ \left| \hat{d}_{k,j}^{\mathrm{s}}-\bar{d}_{k}^{\mathrm{s}}\left( \mathbf{q}_{j}^{\left( 3 \right) ^*} \right) \right| \right\} >\zeta ^{\mathrm{th}}\\
		arg\;\underset{x}{\text{min}}\left\{ \left| \mathcal{D} _k\left( x \right) -\bar{d}_{k}^{\mathrm{s}}\left( \mathbf{q}_{j}^{\left( 3 \right) ^*}\right)\right|  \right\},\quad  \mathrm{otherwise},\\
	\end{cases}\notag
	\\
	&k\notin \mathcal{C}, j\leqslant\tilde{J}.
\end{align}
{\color{blue}
By appropriately choosing the threshold $\zeta ^{\mathrm{th}}$, \eqref{fkj_others} ensures that the ill-conditioned measurements for the $\tilde{J}$ targets in $\left\{ \mathcal{D} _k \right\} _{k\notin \mathcal{C}}$ will not be associated.
Then, the full data association for the $\tilde{J}$ targets is given by 
}
\begin{align}\label{fkj_full}
	\tilde{\mathcal{F}}^* _{\tilde{J}}=\tilde{\mathcal{F}}_{\tilde{J}}^{\left( 3 \right) ^*}\cup \tilde{\mathcal{F}}_{\tilde{J}}^{\mathrm{rest}^*}.
\end{align}
By substituting $\tilde{\mathcal{F}}^*_{\tilde{J}}$ into the subproblems in \eqref{Prob_sub}, we can obtain an accurate location estimation for the $ \tilde{J}$ targets, denoted by $\mathcal{Q} _{\tilde{J}}^{*}=\left\{ \mathbf{q}_{j}^{*}\left| j\leqslant \tilde{J} \right. \right\} $.

Last, the associated bistatic ranges, $\left\{ \mathcal{D} _k\left( \tilde{f}_{k,j}^{*} \right) \right\} _{\tilde{f}_{k,j}^{*}\in \tilde{\mathcal{F}} ^*}$, are eliminated from the measurement set $\mathcal{D} _k$'s.
Then, we can change the combination $\mathcal{C}$ and repeat the above steps until all targets are localized.

\subsubsection{Algorithm Development}\label{subsubsec_Alg_Dev}
\begin{algorithm}[htb]
	\caption{Target Location Estimation Algorithm}
	\label{alg_ML}
	\begin{algorithmic}[1] 
		\Require 
		$\left\lbrace \mathcal{D} _k\right\rbrace _{k=1}^K$, $\left\lbrace \mathbf{a}\right\rbrace _{k=1}^K$, $J$ and the threshold $\zeta ^{\mathrm{th}}$.
		\Ensure 
		The estimation of the targets' location $\mathcal{Q}^{*}$.
		\State \textbf{Initialization}: 
		{\color{blue}Eliminate ill-conditioned measurements which are extremely large or small;}
		Re-index the tAPs and $\mathcal{D} _k$'s;
		Obtain all the possible combination $\mathcal{C}$'s of $c_1$, $c_2$ and $c_3$;
		Define $\mathcal{C}\left( t \right) $ as the $t$-th combination;
		Set $\mathcal{Q}^{*}=\oslash$ and $t=1$. \label{step_initial}
		\ForAll{$t$} \label{step_loopbegin}
		\State 
		$\tilde{J}\gets \min \left\{ \min_{k\in \mathcal{C} \left( t \right)} \left\{ \left| \mathcal{D} _k \right| \right\} ,J-\left| \mathcal{Q} ^* \right| \right\} $;
		\While{$\tilde{J}\geqslant 1$}
		\Comment{\emph{Rough Estimation}}
		\State 
		Solve Problem \eqref{Prob_sub3} over all feasible $\tilde{\mathcal{F}}_{\tilde{J}}^{\left( 3 \right)}$ to obtain 
		\Statex
		$\qquad\;\;\,$ $\tilde{\mathcal{F}}_{\tilde{J}}^{\left( 3 \right) ^*}$ and $\mathcal{Q} ^{\left( 3 \right) ^*}$;
		\State 
		Calculate $\zeta \left( \tilde{\mathcal{F}}_{\tilde{J}}^{\left( 3 \right) ^*},\mathcal{Q} ^{\left( 3 \right) ^*} \right)$ via \eqref{indicator};
		\If{ $\zeta \left( \tilde{\mathcal{F}}_{\tilde{J}}^{\left( 3 \right) ^*},\mathcal{Q} ^{\left( 3 \right) ^*} \right)>\zeta^{\mathrm{th}}$} {$\tilde{J}\gets\tilde{J}-1$}; \label{step_test_tildeJ}
		\Else{ Break};
		\EndIf
		\EndWhile
		\If{$\tilde{J}\geqslant 1$}
		\Comment{\emph{Accurate Estimation}}
		\State 
		Calculate $\tilde{\mathcal{F}}^* _{\tilde{J}}$ via \eqref{fkj_full}; \label{step_cal_tildeF}
		\State
		Solve subproblems in \eqref{Prob_sub} given $\tilde{\mathcal{F}}^* _{\tilde{J}}$ to obtain $\mathcal{Q} _{\tilde{J}}^{*}$;
		\State
		Eliminate $\left\{ \mathcal{D} _k\left( \tilde{f}_{k,j}^{*} \right) \right\} _{\tilde{f}_{k,j}^{*}\in \tilde{\mathcal{F}} ^*}$ from $\mathcal{D} _k$'s; \label{step_elimination}
		\State
		$\mathcal{Q}^{*}\gets \mathcal{Q}^{*}\cup\mathcal{Q} _{\tilde{J}}^{*}$;
		\EndIf
		\If{$\left| \mathcal{Q}^{*}\right| =J$}{ Break;}
		\EndIf
		\EndFor \label{step_loopend}
	\end{algorithmic}
\end{algorithm}
Based on the above discussion, the ML-based algorithm is proposed for target localization, the details of which are summarized in Algorithm \ref{alg_ML}.
Note that with a larger number of range measurements in $\left. \left\{ \mathcal{D} _k \right. \right\} _{k\in \mathcal{C} \left( t \right)}$, more targets are expected to be localized at the $t$-th iteration of Algorithm \ref{alg_ML}.
Therefore, to reduce the total iterations required for estimating all target localizations,
in Step \ref{step_initial}, the tAPs and $\mathcal{D} _k$'s are re-indexed such that $\left| \mathcal{D} _1 \right|\geqslant \left| \mathcal{D} _2 \right|\geqslant \dots \geqslant \left| \mathcal{D} _K \right|$,
and the combination $\mathcal{C}$'s are numbered such that $\sum_{k\in \mathcal{C}\left( t \right)}{k}\leqslant \sum_{k\in \mathcal{C} \left( t+1 \right)}{k}$.

{\color{blue}
It can be noted that the complexity of Algorithm \ref{alg_ML} in solving Problem \eqref{Prob_origin} comes mainly from searching for $\tilde{\mathcal{F}}_{\tilde{J}}^{\left( 3 \right) ^*}$ and solving the $J$ subproblems in \eqref{Prob_sub}.
In specific, the computational complexity for solving each subproblem using the Gauss-Newton algorithm can be derived as $\mathcal{O} \left( r^{\mathrm{GN}}_{\mathrm{iter}}K \right) $, where $r^{\mathrm{GN}}_{\mathrm{iter}}$ represents the number of iterations required to achieve convergence \cite{256302}.
Compared with the exhaustive search method, the proposed Algorithm \ref{alg_ML} can solve Problem \eqref{Prob_origin} with lower complexity while effectively identifying the ill-conditioned measurements.
On one hand, the search overhead of Algorithm \ref{alg_ML} is lower.
Recall that the total number of feasible data association solution $\tilde{\mathcal{F}}$'s over which the exhaustive search method searches is $\prod_{k=2}^K{J!}$.
By contrast, Algorithm \ref{alg_ML} searches over $\frac{\prod_{k\in \mathcal{C}}{\left| \mathcal{D} _k \right|}!}{\hat{J}!\prod_{k\in \mathcal{C}}{\left( \left| \mathcal{D} _k \right|-\hat{J} \right)}!}$ data association solution $\tilde{\mathcal{F}}_{\tilde{J}}^{\left( 3 \right)}$'s to estimate the locations of ${\tilde{J}}$ targets each time.
Although the value of ${\tilde{J}}$ may be determined in an iterative manner, the total number of searches is effectively reduced in most cases.
On the other hand, a data association solution may not satisfy the triangle inequalities in \eqref{triangle_inequal} when ill-conditioned measurements exist.
Hence, the exhaustive search method cannot leverage these inequalities to reduce complexity.
}

\section{Trade-off Analysis and Scenario Extension}\label{Sec_tradeoff_extension}
Based on the proposed sensing scheme for cell-free cooperative ISAC systems, this section analyzes the performance trade-offs achievable for both communication and sensing applications.
Then, through certain modifications, the adaptability of the scheme to general scenarios is enhanced.

\subsection{Performance Trade-offs}
{\color{blue}
In the following, we first discuss the performance trade-offs between communication and sensing applications in the proposed ISAC scheme.
Note that the flexible frame structure defined in the 5G standards introduced in Section \ref{Sec_frame_struc} has paved the way for the trade-offs among a variety of KPIs for each application.
Therefore, some KPI trade-offs are discussed for deeper insights into the cell-free cooperative ISAC systems.

\subsubsection{Trade-offs Between Communication and Sensing}
As mentioned in the system model, the proposed sensing scheme only utilizes the information-bearing signals received at the rAP.
Since the tAPs do not send dedicated sensing reference signals, the sensing functionality does not occupy the time-frequency resources of the existing communication applications.
However, to receive the echo signals from the targets, it is necessary for the APs to be synchronized and employ different slot formats to receive the echoes. 
Therefore, the availability of the sensing functionality depends on the system's flexibility in configuring uplink and downlink transmissions for each AP.
This highlights the advantages of cell-free networks as the hardware platform for ISAC systems, where signal processing and AP transmission configurations are centrally managed by the CPU.
}

\subsubsection{Trade-offs in Communication}
{\color{blue}
As shown in Table \ref{Table_RB}, the 5G standards allow for larger channel bandwidth configurations in the higher frequency band. 
However, due to the higher propagation loss of high-frequency signals, increasing the carrier frequency leads to a reduction in the coverage capability of each AP.
In addition, there is a trade-off in communication networks between the throughput and the delay spread tolerance in transmission. 
On one hand, since the duration of an OFDM symbol is inversely proportional to the SCS $\Delta f^\mu$, more symbols can be transmitted in a frame under the configurations with larger $\mu$, leading to a higher symbol rate. 
On the other hand, the robustness of communication will significantly diminish when the maximum delay spread of the channel exceeds the duration of CP $T^{\text{symb},\mu}_{\text{CP}}$, which decreases as $\Delta f^\mu$ increases.
}

\subsubsection{Trade-offs in Sensing}
{\color{blue}
Similar to the case in communication, there is a trade-off between the bistatic range resolution $\Delta d^{\mathrm{s}}$ and coverage capability.
In the proposed ISAC scheme, $\Delta d^{\mathrm{s}}$ depends on the downlink transmission bandwidth of each AP, thus necessitating a trade-off with maximum detectable bistatic range.
Moreover, each set of bistatic range measurements $\mathcal{D} _k$ is extracted from $M$ received OFDM symbols.
Therefore, a higher achievable refresh rate $r^\mu_{\text{refresh}}$, i.e., the frequency at which the target location is estimated, can be achieved by applying a configuration with larger $\mu$ 
to increase the symbol rate.
Although this adjustment reduces the duration of CP, as will be shown in the simulation results, the sensing performance of the system is not significantly impacted.
}

\subsection{Sensing Scheme for General Scenarios} \label{subsec_SchemeExtension}
{\color{blue}
In the previous derivations and discussions, we assume that all tAPs transmit downlink signals with the same number of active subcarriers and that only one AP is selected as the rAP in each sensing task.
In practical communication networks, however, the UE channel bandwidth can be dynamically configured for the varying transmission capabilities and communication requirements  \cite{3GPP13810101,3GPP13810102}.
Furthermore, the resource allocation flexibility and sensing availability can be enhanced through a cooperative scheme with multiple rAPs.
Consequently, in this subsection, we extend our proposed target localization scheme to accommodate more generalized scenarios.

\subsubsection{Scenarios with Hybrid Bandwidth} \label{subsubsec_HybridBW}
Firstly, we consider the scenarios with \textit{hybrid bandwidth}, where channel bandwidth of the tAPs may be different.
In this case,  we denote the active subcarriers of tAP $k$ and the corresponding bistatic range resolution as $N_{{\mathrm{c}},k}$ and $\Delta d^{\mathrm{s}}_k$, respectively.
}
It is worth noting that the framework introduced in Section \ref{Sec_sens_scheme} remains applicable for target localization, even in cases where the range measurement $\left\lbrace \mathcal{D} _k\right\rbrace $'s exhibit diverse resolutions.
However, to ensure the optimal performance, we propose to modify certain steps in Algorithm \ref{alg_ML}:

\textit{tAP and $\mathcal{D} _k$'s Re-indexing}:
When the target locations are derived from measurements with higher resolutions, the estimation error is expected to reduce.
Therefore, the set of the APs with small $\Delta d_{k}^{\mathrm{s}}$'s should be prioritized to improve the sensing accuracy.
Accordingly, at Step \ref{step_initial} of Algorithm \ref{alg_ML}, the tAPs and $\mathcal{D}_k$'s need to be re-indexed to satisfy the following inequalities for $\forall k\in \left\{ 1,2,\dots ,K-1 \right\} $:
\begin{align}
	&\Delta d_{k}^{\mathrm{s}}\leqslant \Delta d_{k+1}^{\mathrm{s}},
	\\
	&\left| \mathcal{D} _k \right|\geqslant \left| \mathcal{D} _{k+1} \right|, \text{ if } \Delta d_{k}^{\mathrm{s}}=\Delta d_{k+1}^{\mathrm{s}}.
\end{align}

\textit{Thresholds}:
In Algorithm \ref{alg_ML}, a common threshold $\zeta ^{\mathrm{th}}$ is introduced to determine the values of $\tilde{J}$ and $\tilde{f}^*_{k,j}$'s at steps \ref{step_test_tildeJ} and \ref{step_cal_tildeF}, respectively.
Due to variations in the resolutions of $\mathcal{D} _k$'s to be associated, different thresholds are required for rough estimation and accurate estimation in the scenarios with hybrid numerology.
In this case, the threshold at Step \ref{step_test_tildeJ} can be defined as a function $\zeta ^{\mathrm{th}}_{\mathrm{Rou}}$ w.r.t. the set $\left\{ \Delta d_{k}^{\mathrm{s}}\left| k\in \mathcal{C} \left( t \right) \right. \right\} $.
Similarly, when calculating each $\tilde{f}^*_{k,j}$ in \eqref{fkj_others} at Step \ref{step_cal_tildeF}, the threshold $\zeta ^{\mathrm{th}}$ can be replaced by a function $\zeta ^{\mathrm{th}}_{\mathrm{Acc}}$ w.r.t. $\Delta d_{k}^{\mathrm{s}}$.

{\color{blue}
\subsubsection{Scenarios with Multiple rAPs} \label{subsubsec_MultirAP}
When there are $R>1$ rAPs during the duration of the sensing task, we can consider each rAP along with all tAPs as a subsystem for sensing. 
Then, we can obtain $R$ sets of estimation for the locations of the $J$ targets based on the framework in Section \ref{Sec_sens_scheme}.
Denote by $\mathcal{Q} _{r}^{*}=\left\{ \mathbf{q}_{r,j}^{*}\left| j\leqslant J \right. \right\} $ as the set of location estimation generated by the $r$-th subsystem, where $\mathbf{q}_{r,j}^{*}$ is the $r$-th estimated location of target $j$.
Due to the existence of ill-conditioned measurements, we have $\mathbf{q}_{r,j}^{*}=\oslash$ for some $r$.
In this work, we propose to fuse the estimation $\mathcal{Q} _{r}^{*} $'s in result level to improve the localization success rate and accuracy of the cooperative ISAC system.
Specifically, we can associate each estimated location $\mathbf{q}_{r,j}^{*}$ with a target by checking the following inequalities:
\begin{align}
	\left\| \mathbf{q}_{r,j}^{*}-\mathbf{q}_{r\prime,j}^{*} \right\| \leqslant \zeta _{\mathrm{fuse}}^{\mathrm{th}},\; \mbox{if}\,\mathbf{q}_{r,j}^{*},\mathbf{q}_{r\prime,j}^{*}\ne \oslash \,\, \forall j,
\end{align}
where $\zeta _{\mathrm{fuse}}^{\mathrm{th}}$ is a threshold.
Then, a better estimation for each target's location can be obtained as the average of the associated estimated locations.
}

\section{Simulation Results}\label{Sec_simu_resul}
In this section, numerical examples are presented to investigate the performance of the proposed cell-free cooperative ISAC scheme.

\begin{figure}[t]\color{blue}
	\centering  
	\subfigbottomskip=-1pt 
	\subfigcapskip=-5pt 
	\subfigure[LoS path extraction, $M=2$]
	{\label{subFig_LongRange_M2_1}\includegraphics[width=0.32\linewidth]{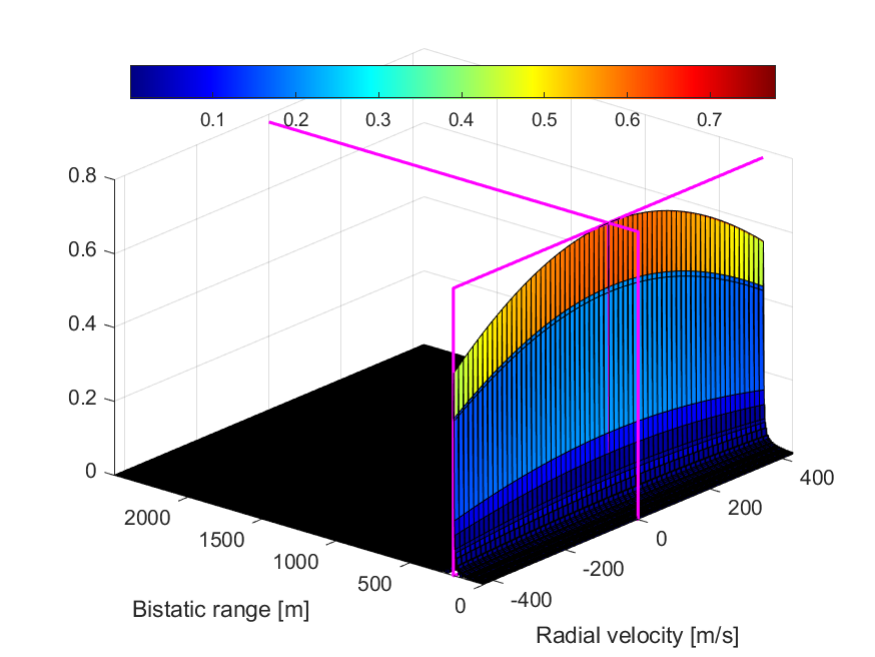}}
	\subfigure[1st scattered path extraction, $M=2$]
	{\label{subFig_LongRange_M2_2}\includegraphics[width=0.32\linewidth]{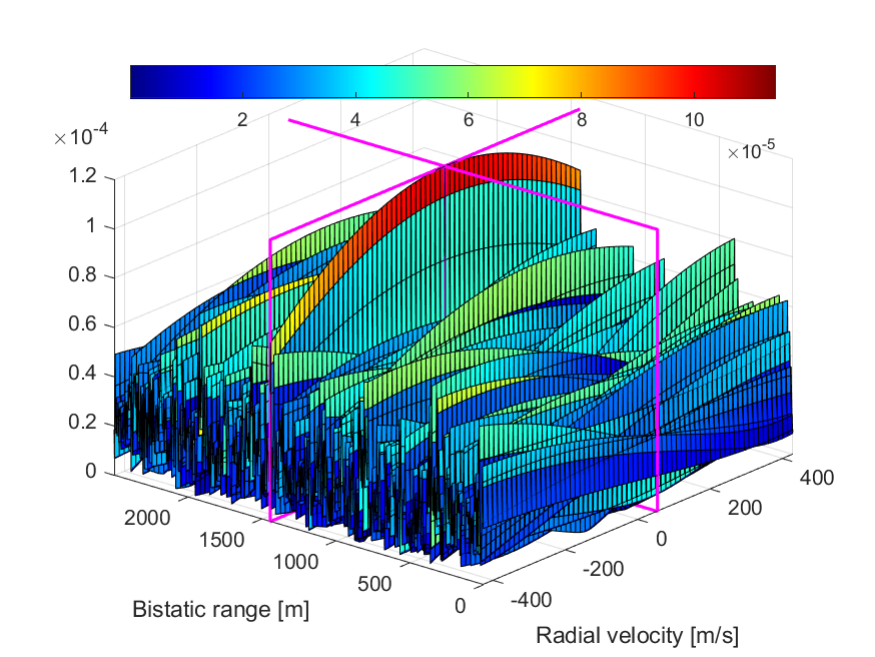}}
	\subfigure[2nd scattered path extraction, $M=2$]
	{\label{subFig_LongRange_M2_3}\includegraphics[width=0.32\linewidth]{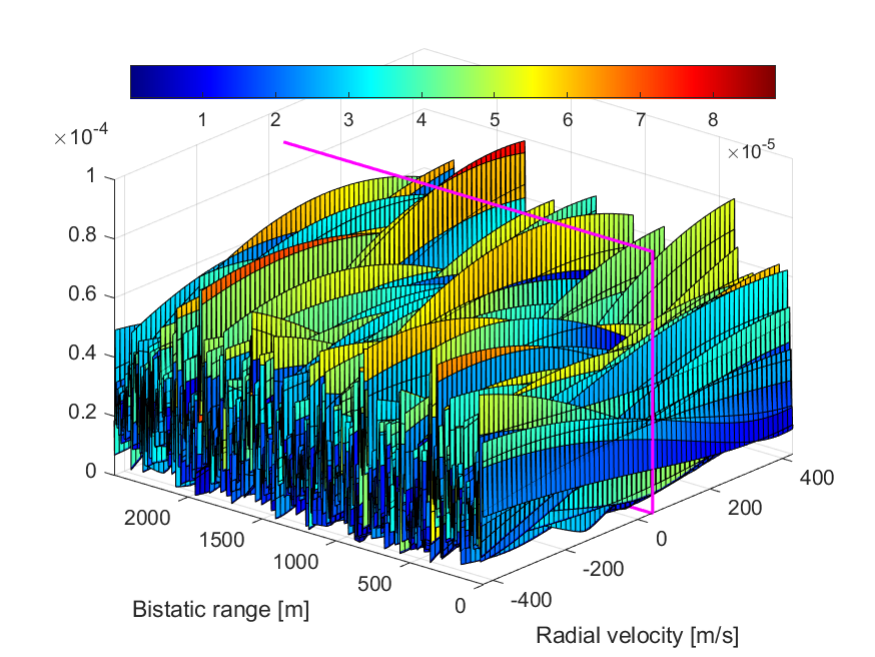}}
		\\
	\subfigure[LoS path extraction, $M=6$]
	{\label{subFig_LongRange_M6_1}\includegraphics[width=0.32\linewidth]{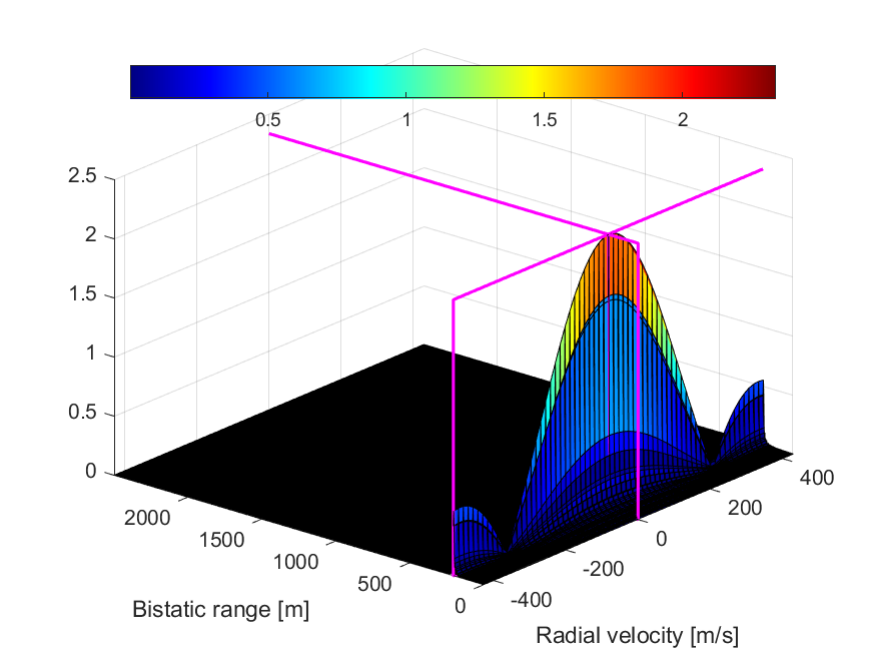}}
	\subfigure[1st scattered path extraction, $M=6$]
	{\label{subFig_LongRange_M6_2}\includegraphics[width=0.32\linewidth]{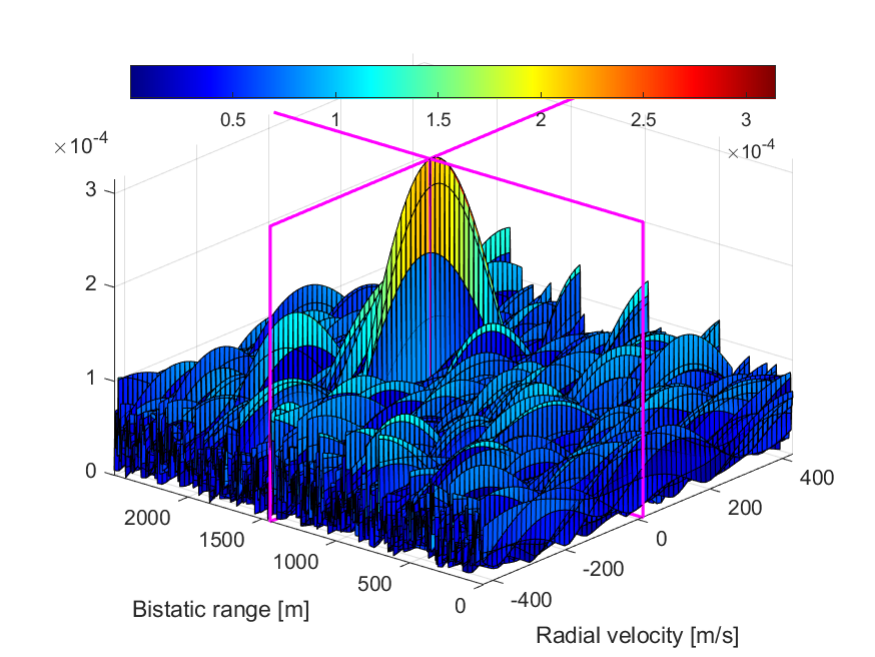}}
	\subfigure[2nd scattered path extraction, $M=6$]
	{\label{subFig_LongRange_M6_3}\includegraphics[width=0.32\linewidth]{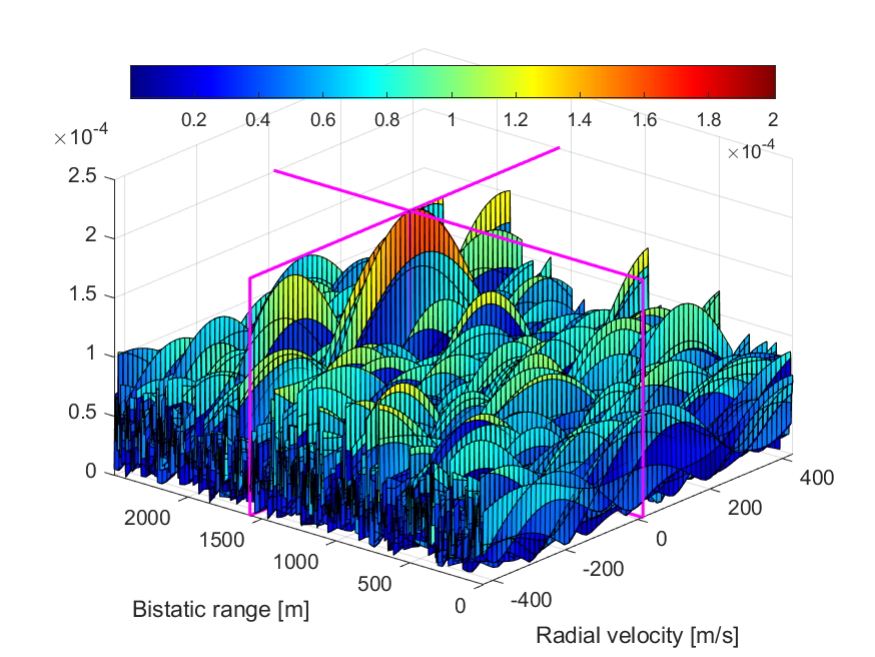}}
		\\
	\subfigure[LoS path extraction, $M=12$]
	{\label{subFig_LongRange_M12_1}\includegraphics[width=0.32\linewidth]{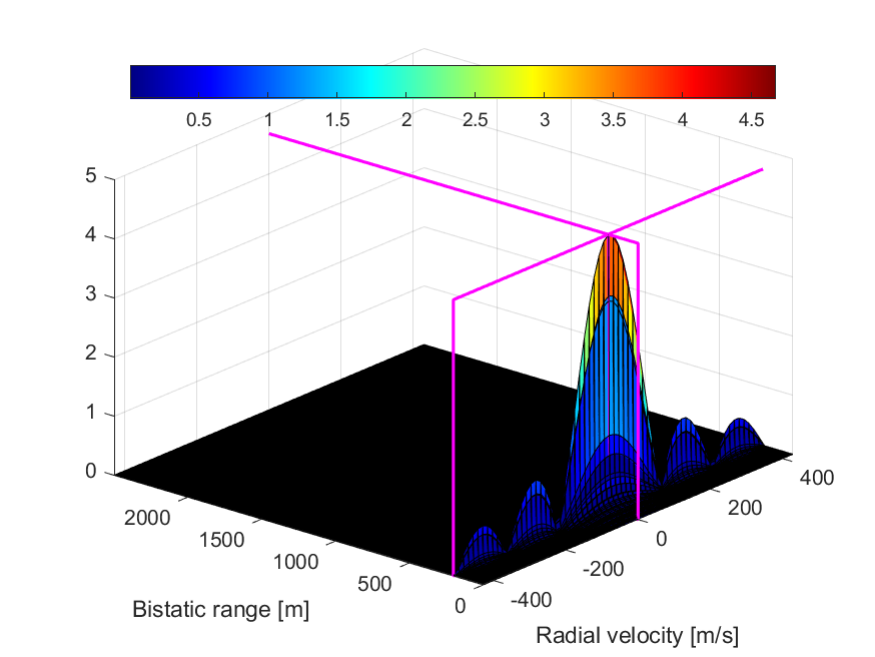}}
	\subfigure[1st scattered path extraction, $M=12$]
	{\label{subFig_LongRange_M12_2}\includegraphics[width=0.32\linewidth]{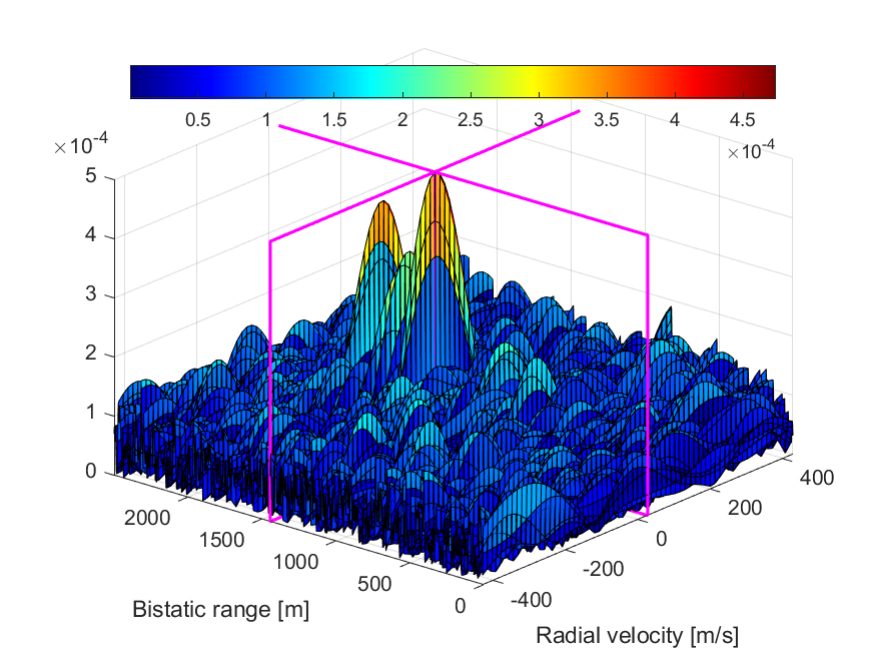}}
	\subfigure[2nd scattered path extraction, $M=12$]
	{\label{subFig_LongRange_M12_3}\includegraphics[width=0.32\linewidth]{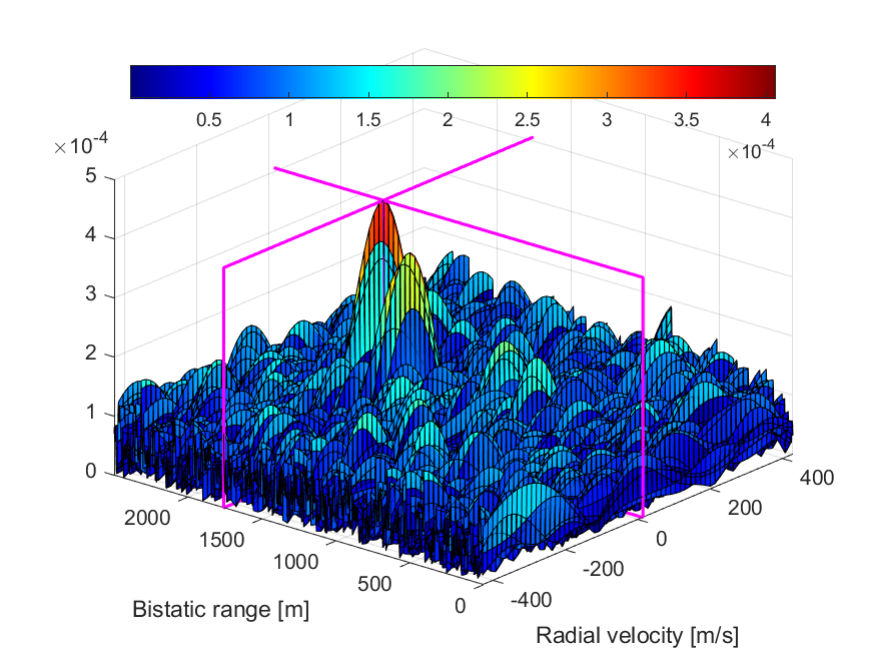}}
		\\
	\subfigure[LoS path extraction, $M=70$]
	{\label{subFig_LongRange_M70_1}\includegraphics[width=0.32\linewidth]{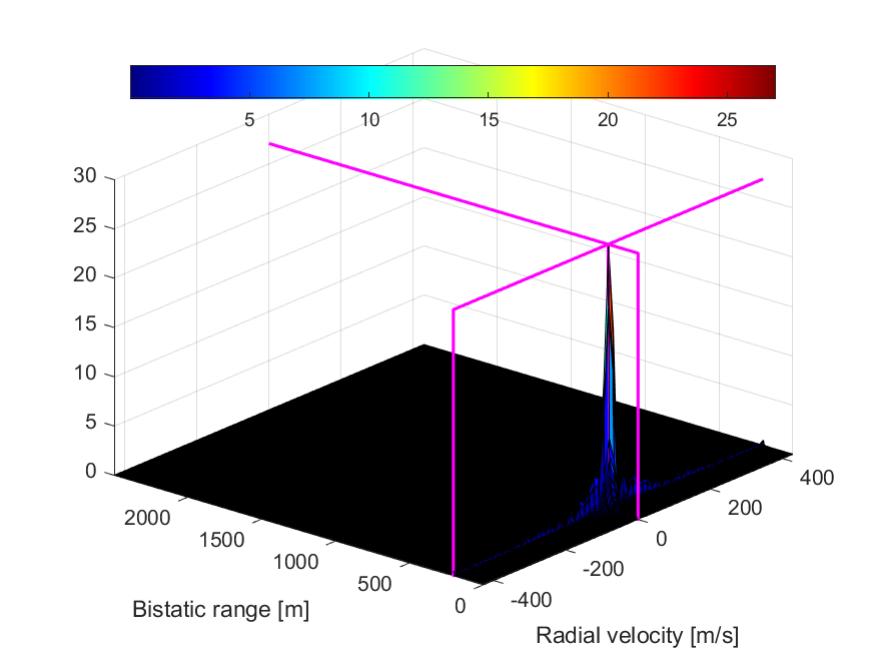}}
	\subfigure[1st scattered path extraction, $M=70$]
	{\label{subFig_LongRange_M70_2}\includegraphics[width=0.32\linewidth]{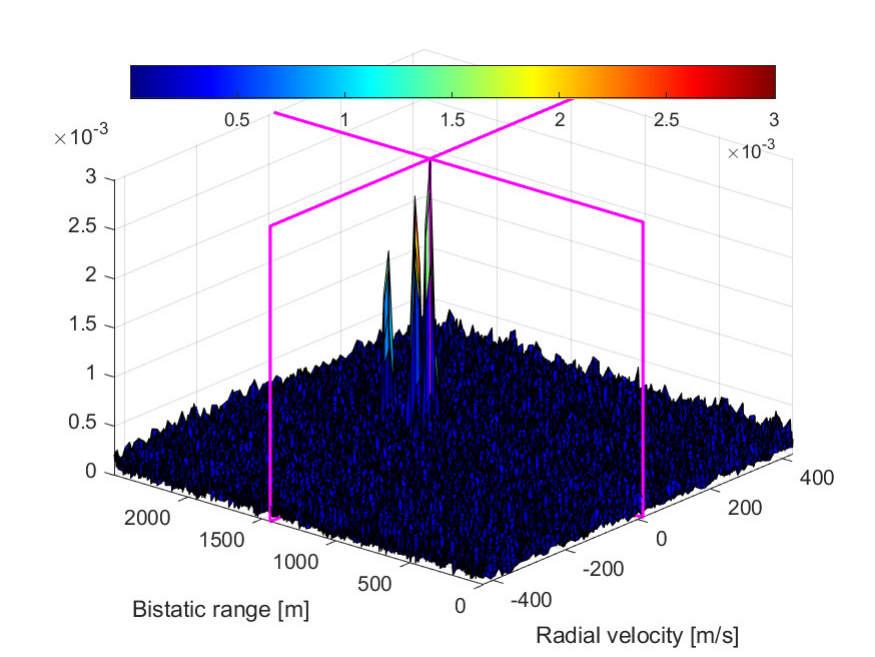}}
	\subfigure[2nd scattered path extraction, $M=70$]
	{\label{subFig_LongRange_M70_3}\includegraphics[width=0.32\linewidth]{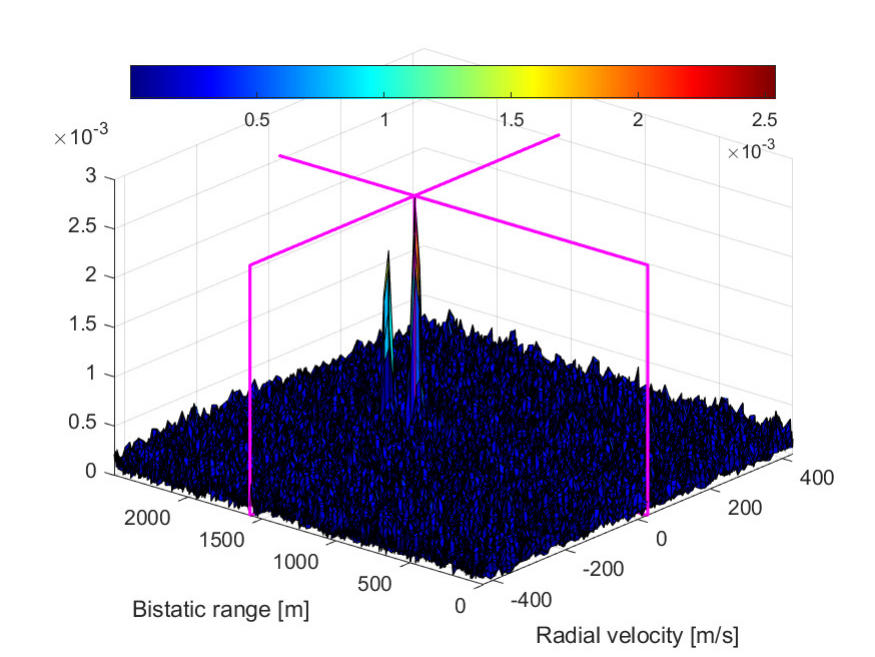}}
	\caption{Examples for the extraction-and-elimination process when the delays of scattered paths exceed the CP duration.}
	\label{Fig_ExtractLongRange}
\end{figure}

{\color{blue}
\subsection{Performance of Range Estimation}
In this subsection, we evaluate the performance for bistatic range estimation of the proposed Algorithm \ref{alg_2DFFT} under the scenario considered in Section \ref{evaluateAlg1}.
The number of targets is set to $J=3$, and the variance of STO and CFO are set to $\sigma _{\mathrm{STO}}^{2}= 10$ ns and $\sigma _{\mathrm{CFO}}^{2}=0.01\Delta f^\mu$, respectively.
Similar to Fig. \ref{Fig_RangeEstiAccu}, only the targets outside the SBZ are accounted for in the results.

\subsubsection{Range Estimation Accuracy}
Fig. \ref{Fig_vsBandwidth} shows the estimation performance under different numerologies and channel bandwidth configurations.
It can be observed that the transmission bandwidth, which determines the resolution $\Delta d^{\mathrm{s}}$ of the 2D-FFT algorithm, leads to significant differences in range estimation accuracy.
The estimation errors for the majority of targets are less than $\Delta d^{\mathrm{s}}/2$, confirming the effectiveness of the proposed Algorithm \ref{alg_2DFFT}.
Additionally, it is noted that within the same channel bandwidth configuration, different SCS settings result in slight variations in estimation error.
This discrepancy arises from the differences in maximum transmission bandwidth as shown in Table \ref{Table_RB}.

\begin{figure}\color{blue}
	\centering
	\includegraphics[width=0.9\linewidth]{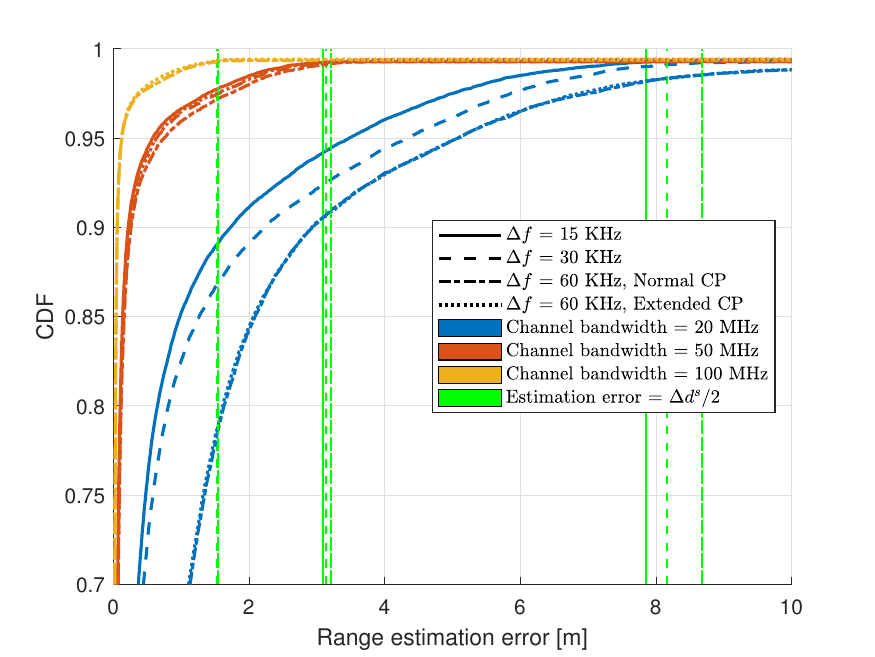}
	\caption{CDF of range estimation error versus numerologies and channel bandwidth.}
	\label{Fig_vsBandwidth}
\end{figure}

\subsubsection{Sensing Availability}
Let us define the success rate of bistatic range estimation as the ratio of the number of measurements with an error less than $\Delta d^{\mathrm{s}}/2$ to the total number of measurements.
To evaluate the sensing availability of the proposed ISAC scheme, Fig. \ref{Fig_SuccessRateVSPt} plots the success rate achieved by Algorithm \ref{alg_2DFFT}.
In Fig. \ref{Fig_SuccessRateVSPt}, the SCS is set to $\Delta f ^\mu =30$ KHz, resulting in a CP duration of 2.3438 $\mu$s.
Hence, when the bistatic range of a target is longer than 703.1 m, the delay of the corresponding scattered path is larger than the CP duration.
However, it can be seen that, with sufficient transmission power $P_{\mathrm{t}}$, Algorithm \ref{alg_2DFFT} maintains a high estimation success rate for long bistatic range targets.
This indicates that the performance of Algorithm \ref{alg_2DFFT} is mainly limited by the echo signal strength, with little impact from the CP duration.

\begin{figure}\color{blue}
	\centering
	\includegraphics[width=0.9\linewidth]{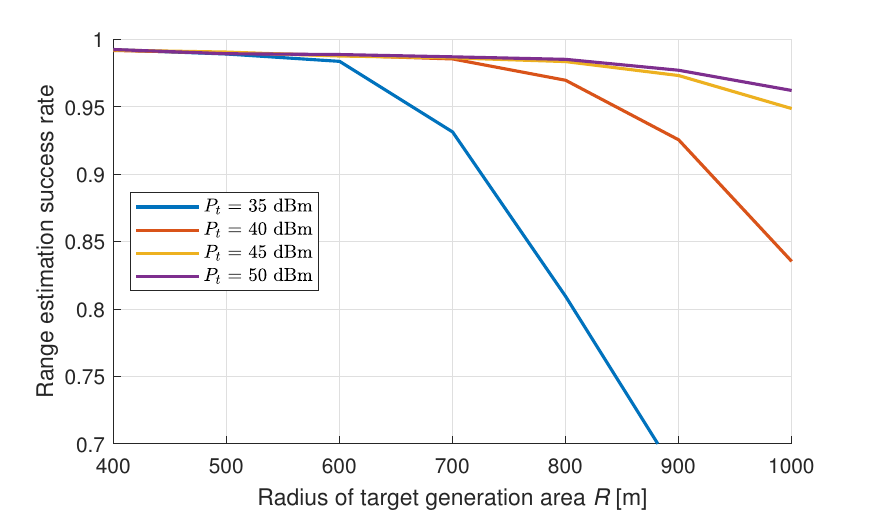}
	\caption{Success rate of bistatic range estimation versus transmit power.}
	\label{Fig_SuccessRateVSPt}
\end{figure}

To better explain this characteristic, we present examples in Fig. \ref{Fig_ExtractLongRange} for the extraction-and-elimination process of Algorithm \ref{alg_2DFFT} when $M=\left[ 2, 6 , 12,70\right] $ symbols are reused for sensing.
In the examples, the bistatic ranges of the targets are 1576.1 m, 1753.1 m and 1440.1 m.
As showm in Fig. \ref{subFig_LongRange_M2_1}-\ref{subFig_LongRange_M2_3}, when there are components with significant delay in the received signal, severe energy leakage occurs in the delay domain of delay-Doppler spectrum $ \mathbf{P}_k $.
However, due to the randomness of the transmitted symbol $s_{k,i,m}$'s, the positions of the spectral peaks caused by energy leakage are also random.
Through the 2D-FFT algorithm, delay components in sensing channels $\left\{ h_{k,i,m}\left| \forall i \right. \right\} $ for $\forall m$ can be extracted simultaneously and then summed.
Therefore, as shown in Fig. \ref{subFig_LongRange_M12_1}-\ref{subFig_LongRange_M70_3}, the delay-Doppler spectrum obtained from the 2D-FFT algorithm still exhibits clear and correctly positioned spectral peaks corresponding to the scattered paths when $M\gg 1$.
Moreover, when $M$ is sufficiently large, the proposed scheme also achieves high resolution in the Doppler dimension, potentially enabling accurate estimation of target velocities.
}

\subsection{Simulation Setup}

\begin{figure}\color{blue}
	\centering
	\includegraphics[width=0.7\linewidth]{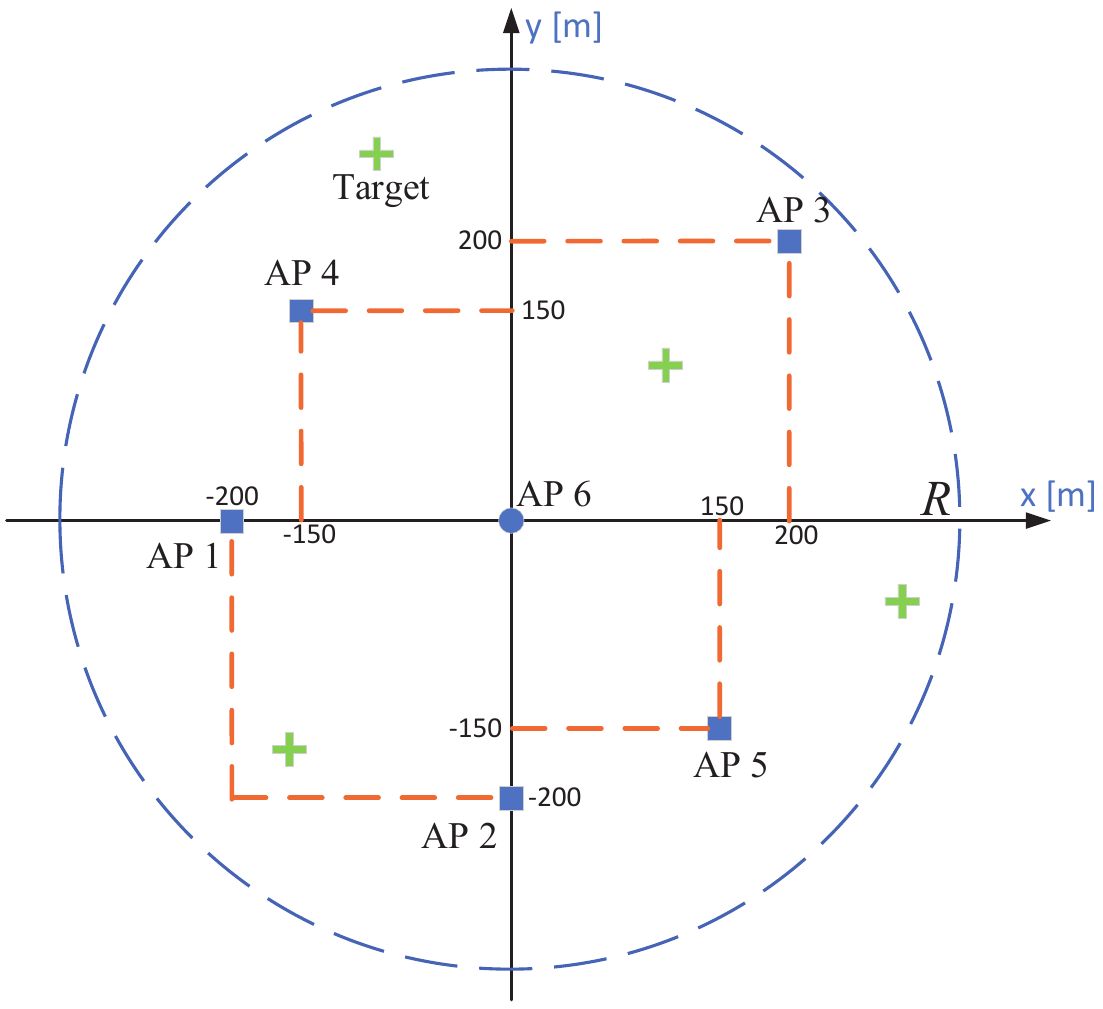}
	\caption{Illustration of the APs' location in simulation.}
	\label{Simu_Model}
\end{figure}

{\color{blue}
In the following, we investigate the overall performance of the proposed cell-free cooperative ISAC scheme.
Fig. \ref{Simu_Model} illustrates the simulated scenario, where five predetermined locations of tAPs (numbered 1-5) are plotted.
The coordinates of the tAPs are set to that of AP 1-$K$ in Fig. \ref{Simu_Model} when simulating a scenario with $K$ tAPs.
Furthermore, each target with a RCS of 1 m$^2$ is randomly and uniformly generated within a circular area with a radius of $R$ m, while the rAP is set as AP 6 at the centre of the circle.
The transmit power of tAP is set to 45 dBm, and the synchronization quality is given by $\sigma _{\mathrm{STO}}^{2}= 10$ ns and $\sigma _{\mathrm{CFO}}^{2}=0.01\Delta f^\mu$.
Unless otherwise stated, the value of $R$ is set to 400.

In the simulation, the bistatic range measurements $\left\lbrace \mathcal{D} _k\right\rbrace _{k=1}^K$ are firstly obtained through Algorithm \ref{alg_2DFFT}, and the target locations are estimated by applying Algorithm \ref{alg_ML}, where the threshold is set to $\zeta ^{\mathrm{th}}=\Delta d^{\mathrm{s}}$.
In addition, we introduce a performance benchmark, which obtains the estimation of target locations by solving Problem \eqref{Prob_origin} given $\mathcal{B} _k=\oslash$ using the exhaustive search method.
The success rate of location estimation is defined as the ratio of the number of correct estimations to the number of all generated targets, 
where a correct estimation is defined as an estimation with an error no greater than $\Delta d^{\mathrm{s}}/2$.
Each simulation result is obtained from $10^4$ independent target generations.
}

%
\subsection{Efficiency of Proposed Algorithm}
\begin{figure}
	\centering
	\includegraphics[width=0.9\linewidth]{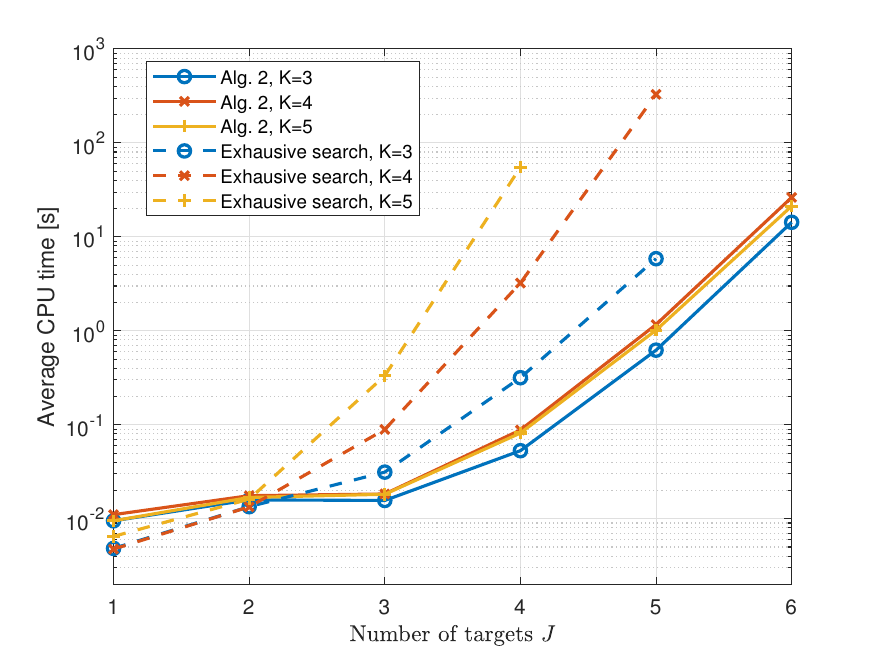}
	\caption{CPU time of Algorithm \ref{alg_ML} and exhaustive search method versus the number of targets $J$.}
	\label{Fig_VStime}
\end{figure}
As discussed in Section \ref{subsubsec_Alg_Dev}, the complexity of the proposed Algorithm \ref{alg_ML} for solving Problem \eqref{Prob_origin} is lower than that of the exhaustive search method. 
In Fig. \ref{Fig_VStime}, we compare the average CPU time of the two algorithms in estimating the target locations under the setup of $K=$ 3, 4, and 5 tAPs.
Note that the y-axis is plotted on a logarithmic scale.
It can be observed that when the number of targets $J$ is less than 2, the CPU time of the exhaustive search method is slightly shorter.
However, when $J$ is relatively large, Algorithm 1 is more efficient. 
Furthermore, the gap in efficiency between the two algorithms increases significantly as $K$ and $J$ increase. 
When $K = 5$ and $J = 4$, more than 99\% computational overhead can be saved by Algorithm \ref{alg_ML}. 
This indicates the superiority of the proposed Algorithm \ref{alg_ML} in multi-target localization tasks within cooperative ISAC systems.

{\color{blue}
\subsection{Localization Performance}
In the following, we evaluate the sensing performance of the cooperative ISAC system.
We denote the proposed sensing information extraction scheme as \textbf{real estimation}, in which the bistatic range measurement $\mathcal{D} _k$'s are obtained thought Algorithm \ref{alg_2DFFT}.
To compare the performance of the proposed Algorithm \ref{alg_ML} and the exhaustive search method in solving Problem \eqref{Prob_origin}, we also consider a baseline scheme with no ill-conditioned measurements, referred to as \textbf{ideal estimation}.
Specifically, this scheme assumes that the factors affecting the accuracy of range estimation in Algorithm \ref{alg_2DFFT} are limited to range resolution, error propagation effects and SBZ, thereby generating the bistatic range measurements in the following manner:
\begin{align}
	\hat{d}_{k,j}^{\mathrm{s}}=\begin{cases}
		\oslash ,\, \mathrm{if}\, d_{k,j}^{\mathrm{s}}\leqslant d_{k}^{\mathrm{b}}+3.5\Delta d^{\mathrm{s}}\\
		d_{k,j}^{\mathrm{s}}+\left( \Delta d^{\mathrm{s}}/2 \right) \cdot \varepsilon \cdot j, \,\mathrm{otherwise}\\
	\end{cases}, \forall k,j,
\end{align}
where $\varepsilon$ is a random variable that follows a normal distribution with zero mean and a variance of 1/9.\footnote{\color{blue}
	Notice that when the target number $J$ is small, the number of ill-conditioned measurements is also small.
	We set the variance to 1/9 because our simulation results show that, with this setup, the success rate of target localization under the \textbf{real estimation} is close to that under the \textbf{ideal estimation} scheme when $J=1$.}
Under this condition, the exhaustive search serves as an upper bound on the localization accuracy of the system, since the optimal data association solution $\mathcal{F} ^*$ for Problem \eqref{Prob_givenF} can be determined based on it.

\begin{figure}\color{blue}
	\centering
	\includegraphics[width=0.95\linewidth]{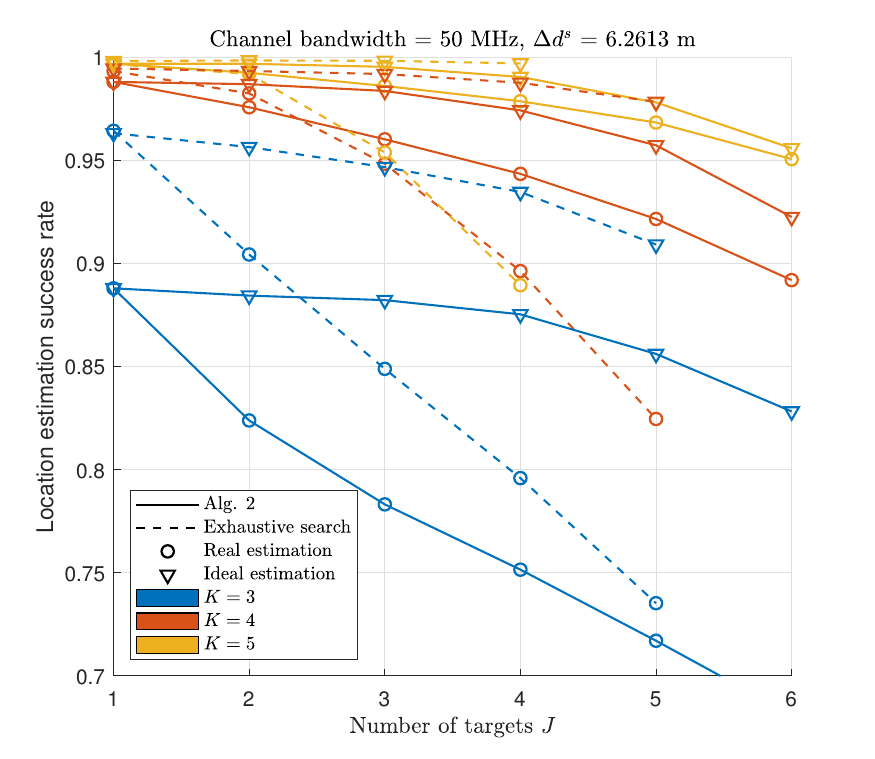}
	\caption{Localization accuracy versus the number of targets $J$ with SCS $\Delta f^\mu = 30$ KHz and channel bandwidth = 50 MHz.}
	\label{Fig_VSaccuracy_f30}
\end{figure}

In Fig. \ref{Fig_VSaccuracy_f30}, we consider a typical case at sub-6G, 
where the SCS is $\Delta f^\mu=30$ KHz and the channel bandwidth of each tAP is 50 MHz,
leading to a bistatic range resolution of $\Delta d^{\mathrm{s}}=6.2613$ m.
The location estimation success rates achieved under the two schemes are plotted in Fig. \ref{Fig_VSaccuracy_f30}.
It is firstly noted that Algorithm \ref{alg_ML} achieves lower accuracy than the exhaustive search method when the number of tAPs $K$ is small.
This is mainly because, when a target is within the SBZs of some tAP-rAP pair, the number of measurements available for estimating its location may be two, i.e., $\left| \left\{ \hat{d}_{k,j}^{\mathrm{s}}\left| \forall k \right. \right\} \right|=2$.
In this case, the exhaustive search algorithm coverages to one of the intersection points of the two bistatic ellipses, which may lead to a correct estimation. 
By contrast, Algorithm \ref{alg_ML} requires at least three bistatic range measurements to generate a location estimation.
However, when the network provides good coverage of the sensing service area ($K\geqslant 5$), the achieved success rates of both algorithms are almost the same.
Secondly, as expected, due to the ill-conditioned measurements, the achieved success rate under \textbf{real estimation} scheme is lower than that under \textbf{ideal estimation} scheme for both algorithms, and this performance gap increases as the number of targets increases.
Interestingly, under \textbf{real estimation} scheme, more ill-conditioned measurements may exist when the number of targets and tAPs is large, causing a sharp decline in the performance of exhaustive methods as shown in Fig. \ref{Fig_VSaccuracy_f30}.
In contrast, Algorithm \ref{alg_ML} demonstrates significantly stronger robustness, highlighting the importance of ill-conditioned measurement elimination, which is not studied in the related works \cite{9724258,10001615,10226276}.
}
Finally, although the success rate performance of our proposed target localization scheme decreases as expected with an increasing number of targets $J$ due to the increase in the probability of mismatching for bistatic range measurements, the decline is not significant as shown in Fig. \ref{Fig_VSaccuracy_f30}.
This validates the capability of the proposed cell-free cooperative ISAC system for multi-target localization.

\begin{figure}\color{blue}
	\centering
	\includegraphics[width=0.95\linewidth]{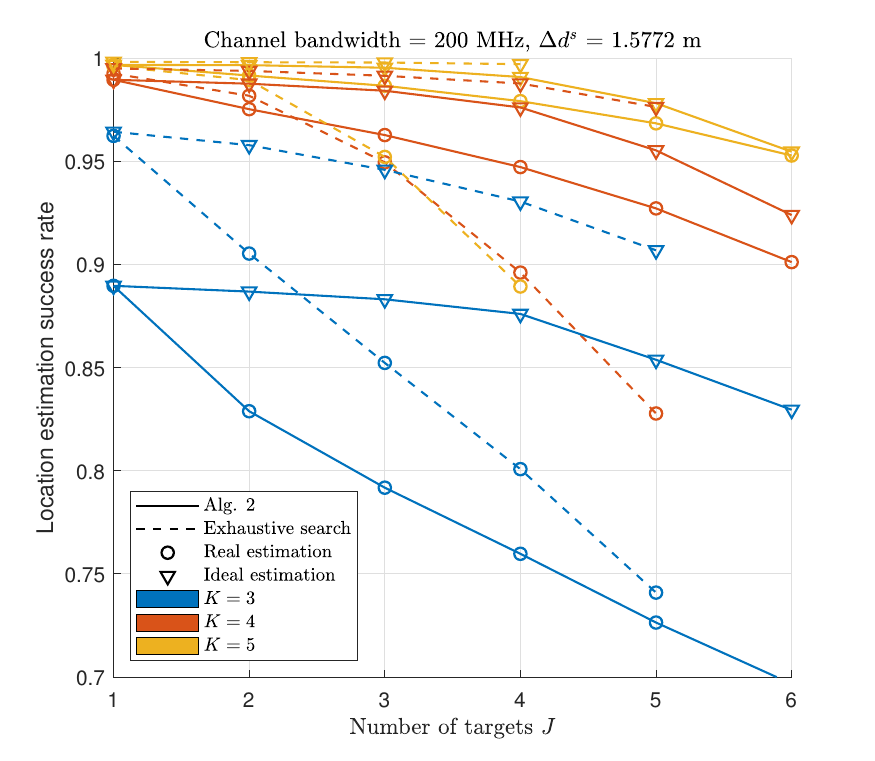}
	\caption{Localization accuracy versus the number of targets $J$ with SCS $\Delta f^\mu = 120$ KHz and channel bandwidth = 200 MHz.}
	\label{Fig_VSaccuracy_f120}
\end{figure}

{\color{blue}
A scenario at FR2 with $\Delta f^\mu=120$ KHz is considered in Fig. \ref{Fig_VSaccuracy_f120}.
The channel bandwidth of each tAP is set to 200 MHz.
Due to the lower coverage capability of high-frequency signals, the radius of the circular area is reduced to $R=$ 100 m and the predetermined locations of tAPs (in m) are set as (-50,0), (0,-50), (50,50), (-35,35) and (35,-35).
It should be noted that in Fig. \ref{Fig_VSaccuracy_f120}, an estimation with an error no greater than $\Delta d^{\mathrm{s}}/2=3.1307$ m is considered correct, while in  Fig. \ref{Fig_VSaccuracy_f30}, an estimation with an error no greater than $\Delta d^{\mathrm{s}}/2=0.7886$ m is deemed correct.
We can observe that the success rate of target localization shown in Fig. \ref{Fig_VSaccuracy_f120} is similar to that in Fig. \ref{Fig_VSaccuracy_f30}, which validates the applicability of our proposed sensing scheme under various system configurations.
Thanks to the improved bistatic range resolution, our proposed scheme exhibits an accuracy of 95\%, ensuring that the localization error in locating $J=6$ targets is within decimeter-level under the scenario with $K=5$ tAPs.
It can be concluded that the proposed sensing scheme possesses good applicability across various transmission configurations.
In addition, from Fig. \ref{Fig_VSaccuracy_f30} and \ref{Fig_VSaccuracy_f120}, the target localization performance of the proposed ISAC scheme can be improved by increasing the number of tAPs $K$.
This highlights the advantages of cooperative sensing networks based on multiple communication infrastructures.
}

\begin{figure}\color{blue}
	\centering
	\includegraphics[width=0.9\linewidth]{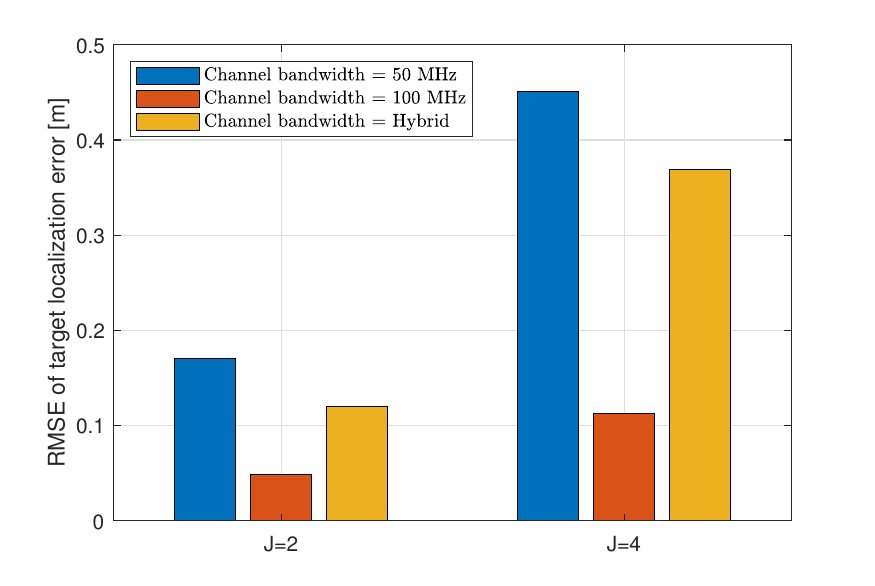}
	\caption{RMSE of localization error under different configurations of channel bandwidth.}
	\label{Fig_RMSEMultiBW}
\end{figure}

{\color{blue}
\subsection{Localization Performance with Hybrid Bandwidth Configuration}
To verify the effectiveness of the modified Algorithm \ref{alg_ML} designed in Section \ref{subsubsec_HybridBW}, we consider the following three schemes of channel bandwidth configuration:
(1) The channel bandwidth of all tAPs is set to 50 MHz;
(2) The channel bandwidth of all tAPs is set to 100 MHz;
and (3) tAP 1-3 are configured with 100 MHz bandwidth while tAP 4-5 are configured with 50 MHz bandwidth.
In the simulation, the thresholds for the modified Algorithm \ref{alg_ML} are set as $\zeta _{\mathrm{Rou}}^{\mathrm{th}}= \max \left\{ \Delta d_{k}^{\mathrm{s}}\left| k\in \mathcal{C} \left( t \right) \right. \right\} $ and $\zeta _{\mathrm{Acc}}^{\mathrm{th}}=\Delta d_{k}^{\mathrm{s}}$.
For scheme (1) and (2), this setup ensures that the performance of the modified Algorithm \ref{alg_ML} is consistent with that of the original algorithm.

Fig. \ref{Fig_RMSEMultiBW} shows the RMSEs of the estimation error achieved by Algorithm \ref{alg_2DFFT} and the modified Algorithm \ref{alg_ML} under the three schemes.
It can be observed that sensing accuracy of the scheme with hybrid bandwidth configuration lies between the other two schemes.
This confirms the compatibility of the modified Algorithm \ref{alg_ML} with general scenarios. 
Furthermore, it indicates that the overall sensing accuracy of the ISAC system can be enhanced by increasing the bandwidth of a portion of the APs.
Therefore, the sensing performance may be enhanced through appropriate frequency domain resource allocation strategies, which suggests a potential direction for future works.

\begin{figure}\color{blue}
	\centering
	\includegraphics[width=0.95\linewidth]{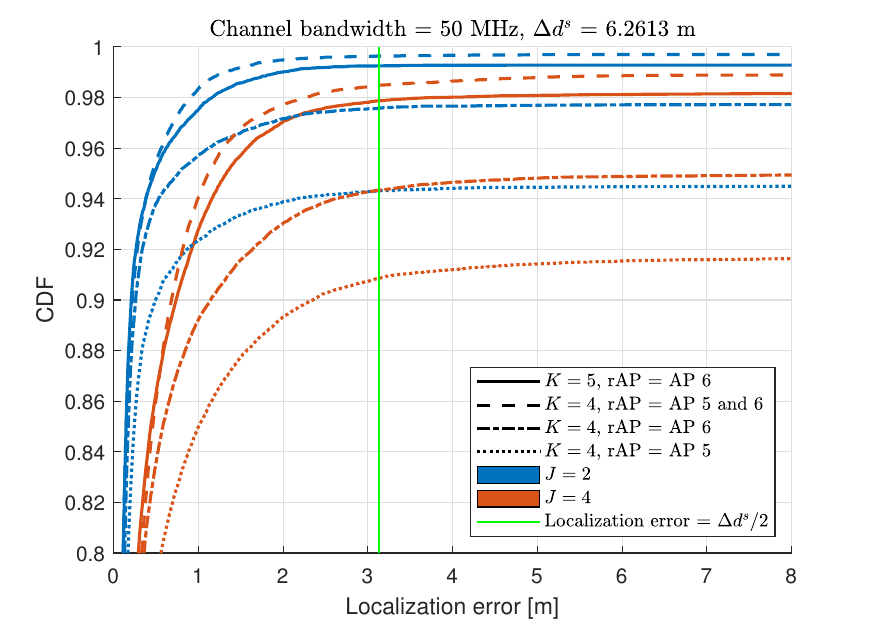}
	\caption{CDF of localization error under different selection of rAP.}
	\label{Fig_MultirAP}
\end{figure}

\subsection{Localization Performance with Multiple rAPs}
Finally, we evaluate the effectiveness of the proposed data fusion scheme for scenarios with multiple rAPs in Section \ref{subsubsec_MultirAP}.
Fig. \ref{Fig_MultirAP} shows the localization error under a scheme with two rAPs (AP 5 and 6 in Fig. \ref{Simu_Model}).
In this case, the six APs are considered to constitute two subsystems for sensing, where AP 1-4 are their common tAPs.
Besides, three schemes with one tAP are evaluated as performance baseline.
The CDFs of localization error under the above schemes are shown in Fig. \ref{Fig_MultirAP}.
As expected, compared with the schemes with only 5 cooperative APs, the 6-AP cooperative sensing schemes achieve higher success rates and accuracy of target localization.
Additionally, in the scheme with two rAPs, each subsystem provides coverage for the SBZs in the other, thereby enhancing the overall network's sensing availability. 
Hence, we can observe that the success rate achieved by the multi-rAP scheme is higher than that of the single-rAP schemes.
}

\section{Conclusion} \label{Sec_conclusion}
In this paper, we proposed a cooperative ISAC framework to integrate the sensing functionality into the cell-free MIMO networks, by leveraging the communication signal transmission of APs and information sharing at the CPU.
{\color{blue}
We designed a two-stage scheme for the bistatic range estimation and the target localization.
In particular, a 2D-FFT-based algorithm was developed to extract the sensing information of the scattered paths while compensating for the STO and CFO between the APs.
Then, an effective algorithm capable of eliminating the ill-conditioned measurements was proposed for joint range data association and location estimation of the targets.
Based on the proposed sensing scheme and 5G NR frame structure, we also discussed the performance trade-offs in practical cooperative ISAC systems.
The sensing scheme was further modified for the general scenarios with hybrid channel bandwidth configuration and that with multiple APs.
Simulation results demonstrated that the proposed sensing scheme achieves high robustness and accuracy in target localization.
}
%
\bibliographystyle{IEEEtran}
\bibliography{IEEEabrv,Refer}

\end{document}